\documentclass[12pt]{article}
\usepackage{latexsym}
\usepackage{amssymb}
\usepackage{graphicx}
\usepackage{multibox}
\usepackage{amsmath,amsfonts}

\topskip \textwidth 480pt

\def\*{\ast}

\def\ve{\varepsilon}

\def\be{\begin{equation}}
\def\ee{\end{equation}}
\def\bqn{\begin{eqnarray}}
\def\eqn{\end{eqnarray}}

\def\theequation{\thesection.\arabic{equation}}

\newsavebox{\ver}
\newsavebox{\verp}

\newsavebox{\gorp}
\newsavebox{\toch}

\newcommand{\bee}{\begin{eqnarray}}
\newcommand{\eee}{\end{eqnarray}}

\newcommand{\ups}{\upsilon}

\begin{document}
\begin{titlepage}
\title{
{\bf String instanton in $AdS_4 \times CP^3$}
~\\
\medskip
\medskip
\medskip
\medskip
\author{Alessandra Cagnazzo\,$^{*\dagger}$, Dmitri
Sorokin\,$^{\dagger}$ and Linus Wulff\,$^{\dagger}$
~\\
~\\
{\it $^*$  Dipartimento di Fisica ``Galileo Galilei",} ~\\
 {\it  Universit\'a degli Studi di Padova} ~\\
~\\
{\it $^\dagger$ INFN, Sezione di Padova,}
~\\
{\it via F. Marzolo 8, 35131 Padova, Italia}}
}

\date{}
\maketitle

\begin{abstract}
We study the string instanton wrapping a non--trivial two--cycle in
$CP^3$ of the type IIA string theory compactified on $AdS_4\times
CP^3$ superspace and find that it has twelve fermionic zero modes
associated with 1/2 of the supersymmetry of the background thus
manifesting that this classical instanton configuration is 1/2 BPS.

\end{abstract}
~

\thispagestyle{empty}
\end{titlepage}
%
\section{Introduction}
M--theory compactified on $AdS_4\times S^7/\mathbb Z_k$ and a corresponding
Type IIA superstring theory compactified on $AdS_4\times CP^3$ are
on the bulk side of the $AdS_4/CFT_3$ holography whose boundary
superconformal Chern--Simons--matter theory is assumed to provide an
effective worldvolume description of a stack of multiple M2--branes
\cite{Schwarz:2004yj,Bagger:2006sk,Bagger:2007jr,Bagger:2007vi,Gustavsson:2007vu,Aharony:2008ug,Benna:2008zy}.
This new $AdS_4/CFT_3$ correspondence shares some features with
the well studied $AdS_5/CFT_4$ correspondence whose bulk theory is
Type IIB superstring theory compactified on $AdS_5 \times S^5$ and the
boundary theory is the superconformal ${\mathcal N}=4$, $D=4$ super
Yang--Mills theory.

However, the two examples of the $AdS/CFT$ correspondence are quite
different. First of all, the Type IIB superstring in $AdS_5 \times
S^5$ is maximally supersymmetric. Its 32 supersymmetries are part of
the superisometry group $PSU(2,2|4)$. On the other hand the
$AdS_4\times S^7/\mathbb Z_k$ solution of $D=11$ supergravity has 32
supersymmetries only for $k=1,2$, while for $k>2$ the theory is
invariant under 24 supersymmetries and so is its ten--dimensional
counterpart, the Type IIA superstring theory in the $AdS_4\times
CP^3$ supergravity background whose superisometries form the
supergroup $OSp(6|4)$
\cite{Nilsson:1984bj,Sorokin:1984ca,Sorokin:1985ap}. As a result,
while the Green--Schwarz action for the $AdS_5 \times S^5$
superstring amounts to a worldsheet sigma--model on the supercoset
space $PSU(2,2|4)/SO(1,4)\times SO(5)$
\cite{Metsaev:1998hf,Bandos:2008ba}, the target superspace of the Green--Schwarz
action of the $AdS_4\times CP^3$ superstring is not a supercoset
space, though it possesses the $OSp(6|4)$ isometry
\cite{Gomis:2008jt,Grassi:2009yj}. Only when the superstring
is extended in $CP^3$, its dynamics can be described by the
$OSp(6|4)/U(3)\times SO(1,3)$ supercoset sigma--model
\cite{Arutyunov:2008if,Stefanski:2008ik,Fre:2008qc,Bonelli:2008us,D'Auria:2008cw}.
It can be obtained from the complete Green--Schwarz action
\cite{Gomis:2008jt} by partially gauge fixing kappa--symmetry in a
way which puts to zero the eight worldsheet fermionic modes
corresponding to the eight broken supersymmetries of the
$AdS_4\times CP^3$ superbackground. Such a gauge fixing is
admissible only when the string moves in $CP^3$. This gauge is,
however not admissible when the string moves entirely in the $AdS_4$
part of the superbackground. One of the consequences of this
peculiar situation is that though the subsector of the $AdS_4\times
CP^3$ superstring theory described by the supercoset sigma--model is
classically integrable \cite{Arutyunov:2008if,Stefanski:2008ik}, the
explicit proof of the classical integrability of the complete
$AdS_4\times CP^3$ superstring still remains an open problem. This
is because the complete $AdS_4\times CP^3$ superspace is not a
supercoset space, and the methods used to prove the integrability of
the $AdS_5\times S^5$ superstring \cite{Bena:2003wd} and of the
supercoset sector of the $AdS_4\times CP^3$ superstring
\cite{Arutyunov:2008if,Stefanski:2008ik} do not apply.

Another interesting peculiarity of the $AdS_4\times CP^3$
superstring, which the $AdS_5\times S^5$ superstring does not have,
is the existence on $CP^3$ of string instantons\footnote{For a
recent review and references on string and brane instantons and
their effects see \emph{e.g.}
\cite{Blumenhagen:2009qh,Bianchi:2009ij}.}. They are formed in the
Wick rotated theory by the string worldsheet wrapping a
topologically non--trivial two--cycle of $CP^3$. This two--cycle is
a $CP^1\simeq S^2$ corresponding to the closed K\"ahler two--form
$J_2$ on $CP^3$. As we shall show, also in the case of the string
instantons on $CP^3$ the consistent gauge fixing of kappa--symmetry
does not allow reducing the string action to the supercoset sigma
model, \emph{i.e.} to eliminate, by using kappa--symmetry, the eight
fermionic modes corresponding to the broken supersymmetries.

The main goal of this paper is to study the superstring instanton on
$CP^3$ and analyze its fermionic zero modes. In the case of branes
moving in a supersymmetric background, their fermionic equations
have solutions which are associated with the Killing spinors of the
background that guarantee its supersymmetries. The number of
physical modes of the worldvolume Dirac operator associated with the
Killing spinors is equal to the number of components of the Killing
spinors which are not annihilated by the kappa--symmetry projector
of the brane. If the background is maximally supersymmetric, one
concludes that the number of dynamical zero modes on the brane are
half the number of supersymmetries, since the rank of the
kappa--symmetry projector is equal to half the number of the maximal
supersymmetries of the background. When the background is not
maximally supersymmetric, as the $AdS_4\times CP^3$ one, the number
of the brane fermion zero modes associated with unbroken
supersymmetries depends on how many of them are not eliminated by
the kappa--symmetry projector. The action of the kappa--symmetry
projector and the corresponding number of fermionic zero modes
depends on how the given brane configuration is embedded into target
space. In the cases with less supersymmetries the worldvolume Dirac
equation may, in general, also have solutions which are not
associated with unbroken supersymmetries, but with the broken ones.
Thus the analysis of the brane fermionic modes in less
supersymmetric backgrounds should be made case by case (see
\emph{e.g.} \cite{Bergshoeff:2005yp} for a more detailed discussion of this
point).

As we shall see, in the $AdS_4\times CP^3$ case the string instanton
on $CP^3$ has twelve zero modes all of which are associated with the
supersymmetries of the background and there are no zero modes of the
fermions corresponding to the supersymmetries broken by $AdS_4\times
CP^3$. So the instanton under consideration is 1/2 BPS. It is
interesting that these twelve zero modes are divided into eight and
four ones which have different geometrical and physical meaning. The
eight massive fermionic zero modes are four copies of the
two--component Killing spinor on $S^2$ and the four other fermionic
modes are two copies of massless chiral and anti-chiral fermion on
$S^2$ electrically coupled to the electromagnetic potential created
on $S^2$ by a monopole placed in the center of $S^2$. The monopole
potential arises as part of the $CP^3$ spin connection pulled--back
on the instanton $S^2$. This, at least formally and remotely,
reminds us the peculiarity of the presence of different, light and
heavy, physical worldsheet degrees of freedom in a Penrose limit of
the $AdS_4\times CP^3$ superstring
\cite{Nishioka:2008gz,Grignani:2008is,Zarembo:2009au,Sundin:2009zu}.

In M--theory compactified on $AdS_4\times S^7/\mathbb Z_k$, the
counterpart of the string instanton considered in this paper is an
Euclidean M2--brane that wraps the non--trivial 3--cycle
$S^3/\mathbb Z_k$ (for $k>1$) inside $S^7/\mathbb Z_k$.

The presence of the string instanton and its fermionic zero modes
may generate non--perturbative corrections to the string effective
action, which may affect its properties and if so should be taken
into account in studying, \emph{e.g.} the $AdS_4/CFT_3$
correspondence.  The instantons may, perhaps, contribute to the
worldsheet S--matrix and/or to energies of a semiclassical string.
To study these effects one needs to find a way of merging the
instanton and Minkowski solutions, such as spinning strings or BMN
geodesics. In addition, in the presence of the instanton fermionic
zero modes, the worldsheet correlator, to be non--zero, should
contain a number of fermion insertions\footnote{We are thankful to
Konstantin Zarembo for these comments.}.

The paper is organized as follows. In Section \ref{quada} we review
the form of the string action in the $AdS_4\times CP^3$
superbackground and truncate it to the second order in fermions. In
Section \ref{bi} we describe the bosonic part of the string
instanton solution and in Section \ref{fmodes} we study its
fermionic zero modes. Section \ref{summary} contains a summary and
brief discussion of a possible relation of the string instanton to
some features of the $AdS_4/CFT_3$ corespondence. Appendix A
contains a description of our conventions and notation and of the
geometry of the $AdS_4\times CP^3$ superspace. In Appendix B we give
the explicit form of the $CP^3$ Fubini--Study metric, vielbeins and
connection which were used for the analysis of the string fermion
equations.

\section{$AdS_4\times CP^3$ superstring action up to the quadratic order in
fermions}\label{quada}

To simplify the study of the fermionic zero modes of the string
instanton we reduce the complete superstring action of
\cite{Gomis:2008jt} to the quadratic order in fermions (though, the
solutions we find satisfy the complete non--linear equations of
motion to all orders in fermions). Alternatively, one can use the
quadratic type IIA superstring action derived in
\cite{Tseytlin:1996hs,Cvetic:1999zs} for a generic superbackground
and substitute into it the values of the supergravity fields
corresponding to the $AdS_4\times CP^3$ background. As a consistency
check we have performed the reduction of both of the actions. We
shall see that they give the same result upon a redefinition of
bosonic $CP^3$ coordinates of the target--superspace of
\cite{Gomis:2008jt}. This redefinition is required due to a
particular parametrization used in \cite{Gomis:2008jt} to construct
an explicit form of the $AdS_4\times CP^3$ supergeometry.

The Green--Schwarz superstring action in a generic type IIA
supergravity background has the well known form
\begin{equation}\label{cordaA}
S = -\frac{1}{4\pi\alpha'}\,\int d^2\xi\, \sqrt {-h}\, h^{IJ}\,
{\cal E}_I{}^A {\cal E}_J{}^B \eta_{AB}
-\frac{1}{2\pi\alpha'}\,\int  B_2\,,
\end{equation}
where $\xi^I$  $(I,J=0,1)$ are the worldsheet coordinates,
$h_{IJ}(\xi)$ is an intrinsic worldsheet metric, ${\mathcal
E_J{}^A}$ are worldsheet pullbacks of target superspace vector
supervielbeins and $B_2$ is the pull--back of the NS--NS 2--form.

The kappa--symmetry transformations  of the worldsheet fields
$Z^{\mathcal M}(\xi)=(X^M(\xi),\Theta^{\underline \alpha}(\xi))$
which leave the superstring action (\ref{cordaA}) invariant
(provided the superbackground obeys the superspace supergravity
constraints) are\footnote{The kappa--variations of $Z^{\mathcal M}$
should be accompanied by a kappa--variation of the worldsheet metric
$h_{IJ}$ whose explicit form the reader may find in
\cite{Grassi:2009yj}, eq. (4.21).}
\begin{equation}\label{kappastring}
\delta_\kappa Z^{\mathcal M}\,{\mathcal E}_{\mathcal M}{}^{\underline \alpha}=
\frac{1}{2}(1+\Gamma)^{\underline \alpha}_{~\underline\beta}\,
\kappa^{\underline\beta}(\xi),\qquad {\underline \alpha}=1,\cdots, 32
\ee
\be\label{kA}
\hskip-2.5cm\delta_\kappa Z^{\mathcal M}\,{\mathcal E}_{\mathcal M}{}^A=0,
\qquad   A=0,1,\cdots,9
\end{equation}
where $\kappa^{\underline\alpha}(\xi)$ is a 32--component spinor
parameter, $\frac{1}{2}(1+\Gamma)^{\underline
\alpha}_{~\underline\beta}$ is a spinor projection matrix with
\be\label{gbs}
\Gamma=\frac{1}{2\,\sqrt{-\det{G_{IJ}}}}\,\varepsilon^{IJ}\,{\mathcal
E}_{I}{}^A\,{\mathcal E}_{J}{}^B\,\Gamma_{AB}\,\Gamma_{11}, \qquad
\Gamma^2=1\,,
\ee
where $G_{IJ}={\mathcal E}_I{}^A\,{\mathcal E}_J{}^B\,\eta_{AB}$ is
the induced metric on the worldsheet. The explicit form of the supervielbeins
${\mathcal E}^A(Z)$ and the NS--NS 2--form $B_2$ which describe the
geometry of the $AdS_4\times CP^3$ superspace are given in Appendix A (see also
\cite{Gomis:2008jt,Grassi:2009yj}).

Up to second order in fermions the supervielbeins and the $B$-field
have the following form
\if{0}
\begin{equation}
e^{\frac{2}{3}\phi}=e^{\frac{2}{3}\phi_0}(1-\frac{2}{R}\ups\gamma^5\ups)
\qquad\lambda^{\underline\alpha}=e^{\frac{1}{3}\phi_0}\frac{2i}{R}(\gamma^5\ups)^{\underline\alpha}
\end{equation}
\fi
\begin{eqnarray}\label{E2}
\mathcal E^a&=&e^{\frac{1}{3}\phi_0}\Big(e^a(1-\frac{1}{R}\ups\gamma^5\ups)+i\Theta\gamma^a\mathcal D\Theta\Big),\nonumber\\
\mathcal E^{a'}&=&e^{\frac{1}{3}\phi_0}\Big(e^{a'}(1-\frac{1}{R}\ups\gamma^5\ups)
+i\vartheta\gamma^{a'}\gamma^5\mathcal D\vartheta
+2i\ups\gamma^{a'}\gamma^5\mathcal D\vartheta\Big),\\
\mathcal E^{\underline\alpha}&=&e^{\frac{1}{6}\phi_0}(\mathcal D\Theta)^{\underline\alpha}\,,\nonumber
\end{eqnarray}
\bee\label{B22}
B_2&=&e^{\frac{2}{3}\phi_0}\big(ie^A\,\Theta\Gamma_A\Gamma_{11}\mathcal D\Theta-\frac{1}{R}e^Be^A\,\Theta\Gamma_{AB}\gamma_7\ups\big)
\,,
\eee
where $e^{\frac{2}{3}\phi_0}=\frac{R}{kl_p}$ is the vacuum
expectation value of the dilaton, $R$ is the radius of the $S^7$
sphere whose base is $CP^3$, $l_p$ is the eleven-dimensional Planck
length related to the string tension as follows
$l_p=e^{\frac{1}{3}\phi_0}\sqrt{\alpha'}$ and $k$ is the
Chern--Simons level related to the units of $F_4$ and $F_2$
Ramond--Ramond flux which support the $AdS_4\times CP^3$ solution of
Type IIA supergravity. The contribution of the RR fluxes manifests
itself in the presence of projectors $\mathcal P_6$ and $\mathcal
P_2$ in the string action (see Appendix A.5). They split the
32--component fermionic variable $\Theta^{\underline\alpha}$  into
the 24--component spinors $\vartheta^{\alpha a'}$
($\alpha=1,\ldots,4$; $a'=1,\ldots, 6$) which correspond to the 24
supersymmetries of the $AdS_4\times CP^3$ solution and the
8--component spinors $\upsilon^{\alpha q}$ $(q=1,2)$ which
correspond to the broken supersymmetries. The index $\alpha$ is a
spinor index of $AdS_4$ (see Appendix A for more details). The
covariant derivative $\mathcal D\Theta$ is defined as follows
\begin{equation}
\mathcal D\Theta=
\left\{\begin{array}{l}
\mathcal D\ups=(d+\frac{i}{R}e^a\gamma^5\gamma_a-\frac{1}{4}\omega^{ab}\gamma_{ab}-2iA\gamma_7)\ups\\
\mathcal D\vartheta=\mathcal P_6(d+\frac{i}{R}e^a\gamma^5\gamma_a+\frac{i}{R}e^{a'}\gamma_{a'}-\frac{1}{4}\omega^{ab}\gamma_{ab}-\frac{1}{4}\omega^{a'b'}\gamma_{a'b'})\vartheta\\
\end{array}\right.\,,
\end{equation}
where $e^a(x)$, $\omega^{ab}(x)$ and $\gamma^a,\gamma^5$ are,
respectively the vielbein, connection and Dirac--matrices of $AdS_4$
of radius $R/2$. $e^{a'}(y)$ and $\omega^{a'b'}(y)$ are,
respectively, the vielbein and connection on $CP^3$ of radius $R$
and $\gamma^{a'},\gamma^7$ are $8\times 8$ gamma--matrices of
$Spin(6)$. $A(y)=\frac{1}{8}\,\omega^{a'b'}\,J_{a'b'}$ is the RR
one-form potential whose field strength is the K\"ahler form on
$CP^3$, $dA=\frac{1}{R^2}e^{a'}e^{b'}J_{a'b'}$. See Appendix A for
more details regarding the notation and conventions.

Substituting the expressions for the vielbeins \eqref{E2} and the
NS--NS two--form \eqref{B22} into the action \eqref{cordaA} and
keeping only terms up to quadratic order in fermions we get
the following action
\bee\label{action2}
S&=&-\frac{e^{\frac{2}{3}\phi_0}}{4\pi\alpha'}\,\int d^2\xi\, \sqrt
{-h}\, h^{IJ}\,\left( {e}_{I}{}^{a} {e}_{J}{}^{b}
\eta_{ab}+{e}_{I}{}^{a'} {e}_{J}{}^{b'}\,\delta_{a'b'}\right)\nonumber\\
&-&
\frac{e^{\frac{2}{3}\phi_0}}{2\pi\alpha'}\,\int
d^2\xi\,\Theta(\sqrt{-h}\,h^{IJ}-\varepsilon^{IJ}\Gamma_{11})\big[i\,e_I{}^A\Gamma_A\nabla_J\Theta
-\frac{1}{R}e_I{}^Ae_J{}^B\Gamma_A\mathcal P_6\gamma^5\Gamma_B\Theta\big]\nonumber\\
&+&\frac{e^{\frac{2}{3}\phi_0}}{2\pi\alpha'}\,\int d^2\xi\,\sqrt
{-h}\,h^{IJ}\,e_I{}^{a'}\,\nabla_J(i\,\Theta{\mathcal P}_6\Gamma_{a'}{\mathcal P}_2\Theta)\,,
\eee
where $\nabla\Theta=(d-\frac{1}{4}\,\omega^{AB}\,\Gamma_{AB})\Theta$ is the
worldsheet pullback of the conventional $AdS_4\times CP^3$ covariant
derivative.

The first two lines of this action coincide with the action which
one gets by reducing to $AdS_4\times CP^3$ the quadratic
Green--Schwarz action in a generic type IIA superbackground
\cite{Tseytlin:1996hs,Cvetic:1999zs}.  The last term in the action
\eqref{action2}
appeared because of our choice of parametrization of the
$AdS_4\times CP^3$ superspace which allowed us to write its geometry
in the simplest form. It is not hard to see that the last term in
\eqref{action2} can be canceled (modulo higher order terms in
fermions) by making the following shift of the bosonic coordinates
$y^{m'}$ of $CP^{3}$
\be\label{y}
y^{m'}=\hat y^{m'}+i\Theta{\mathcal P}_6\Gamma^{a'}{\mathcal
P}_2\Theta\,e_{a'}{}^{m'}(\hat y).
\ee
After this field redefinition the two forms of the string action
become equivalent.

To study the string instantons we should perform a Wick rotation of
the worldsheet and the target space in the action \eqref{action2} to
Euclidean signature. The Wick rotation basically consists in
replacing $\sqrt{-h}$ and $\sqrt{-G}$, respectively with $\sqrt{h}$
and $\sqrt{G}$, replacing $\varepsilon^{IJ}$  with
$-i\,\varepsilon^{IJ}$ and taking into account that the fermions
$\Theta$ become complex spinors, since there are no Majorana spinors
in ten-dimensional Euclidean space. However, the complex conjugate
spinors do not appear in the Wick rotated action and, hence, the
number of the fermionic degrees of freedom formally remains the same
as before the Wick rotation. Note also that the Euclidean $\gamma^5$
is defined as $\gamma^5=\gamma^1\,\gamma^2\,\gamma^3\,\gamma^4$,
where $\gamma^4$ is the Wick rotated $\gamma^0$. So $(\gamma^5)^2=1$
as in the case of Minkowski signature.

Thus, after the redefinition \eqref{y} and the Wick rotation the
action takes the following form
\bee\label{wickra}
S_E&=&\frac{e^{\frac{2}{3}\phi_0}}{4\pi\alpha'}\,\int d^2\xi\, \sqrt
{h}\, h^{IJ}\,\left( {e}_{I}{}^{a} {e}_{J}{}^{b}
\delta_{ab}+{e}_{I}{}^{a'} {e}_{J}{}^{b'}\,\delta_{a'b'}\right)\\
&+&
\frac{e^{\frac{2}{3}\phi_0}}{2\pi\alpha'}\,\int
d^2\xi\,\Theta(\sqrt {h}\,h^{IJ}+i\varepsilon^{IJ}\Gamma_{11})\big[i\,e_I{}^A\Gamma_A\nabla_J\Theta
-\frac{1}{R}e_I{}^Ae_J{}^B\Gamma_A\mathcal P_6\gamma^5\Gamma_B\Theta\big]
\nonumber
\eee
and the kappa--symmetry matrix $\Gamma$ gets replaced by
\be\label{gbse}
\Gamma=-\frac{i}{2\,\sqrt{\det{G_{IJ}}}}\,\varepsilon^{IJ}\,{\mathcal
E}_{I}{}^A\,{\mathcal E}_{J}{}^B\,\Gamma_{AB}\,\Gamma_{11}, \qquad
\Gamma^2=1\,.
\ee

\section{String instanton wrapping a two--sphere inside
$CP^3$}\label{bi} We are interested in a string whose worldsheet
wraps a topologically non--trivial two--cycle inside $CP^3$ and thus
is a stringy counterpart of the instantons of two--dimensional
$CP^N$ sigma--models\footnote{The instanton solution in the $O(3)$
(or $CP^1$) sigma--model was first found in \cite{Polyakov:1975yp}
and then generalized to the case of the $CP^n$ sigma--models in
\cite{Golo:1978de,Golo:1978dd,D'Adda:1978uc}. The instanton solution in the supersymmetric
$CP^1$ sigma--model was first discussed in \cite{DiVecchia:1977bs}.
See
\cite{Novikov:1984ac,Perelomov:1987va,Perelomov:1989im} for a review and references on this subject.}. To be
topologically non--trivial this two--cycle should have a non--zero
pull--back on its worldsheet of the K\"ahler two--form
$J_2=\frac{1}{2}\,e^{b'}\,e^{a'}\,J_{a'b'}$ of $CP^3$. Such a
two--cycle is a $CP^1\simeq S^2$ subspace of $CP^3$. To identify it,
it is convenient to consider the form of the Fubini--Study metric on
$CP^3$ given in \cite{Pope:1984bd}
\begin{equation}\label{metric}
ds^2=
R^2\big[\frac{1}{4}\big(d\theta^2+\sin^2\theta(d\varphi+\frac{1}{2}\sin^2\alpha\,\sigma_3)^2\big)
+\sin^2\frac{\theta}{2}\,d\alpha^2
+\frac{1}{4}\sin^2\frac{\theta}{2}\,\sin^2\alpha(\sigma_1^2+\sigma_2^2+\cos^2\alpha\,\sigma_3^2)\big]\,,
\end{equation}
where $0\leq\theta\leq\pi$, $0\leq\varphi\leq2\pi$ and
$0\leq\alpha\leq\frac{\pi}{2}$, and $\sigma_1,\sigma_2,\sigma_3$
are three left-invariant one-forms on $SU(2)$ obeying
$d\sigma_1=-\sigma_2\sigma_3$ etc. (see Appendix B for more
details). Notice that with this choice of the $CP^3$
 coordinates, $\theta$ and $\varphi$ parameterize a two--sphere of
radius $\frac{R}{2}$. This two--sphere is topologically non--trivial
and associated to the K\"ahler form on $CP^3$. The string instanton
wraps this sphere. For instance, if it wraps the sphere once
$\theta$ and $\varphi$ can be identified with the string worldsheet
coordinates, while all other $CP^3$ as well as $AdS_4$ coordinates
are worldsheet constants in this case. Thus the pullback on the
string instanton of the metric \eqref{metric} of $CP^3$ (of radius
$R$) is the metric of the sphere of radius $R/2$
\be\label{s2metric}
ds^2=\frac{R^2}{4}\,(d\theta^2+\sin^2\theta\,d\varphi^2\,)\,.
\ee
In this coordinate system the $S^2$ vielbein $e^i$ and the spin
connection $\omega^{ij}_{S^2}$ $(i,j=1,2)$ can be chosen in the form
\be\label{os2}
e^1=\frac{R}{2}d\theta\,,\qquad e^2=\frac{R}{2}
\sin\theta\,d\varphi\,,\qquad
\omega^{12}_{S^2}=\cos\theta d\varphi\,,
\ee
and the $S^2$ curvature is
\be\label{curvature}
R^{ij}=d\omega^{ij}_{S^2}=\frac{4}{R^2}\,e^i\,e^{j}\,.
\ee

\subsection{Bosonic part of the instanton solution}\label{bipart}
The bosonic part of the Wick rotated string action \eqref{wickra} is
\be
S_E=\frac{T}{2}\,\int d^2\xi\,\sqrt{h}\, h^{IJ}\,e_I{}^ie_J{}^j\,\delta_{ij}\,,
\ee
where $T=\frac{e^{\frac{2}{3}\phi_0}}{2\pi\alpha'}$ and $e^i$ are
the vielbeins on $S^2$. To discuss the instanton solution of this
$CP^1$ sigma model it is convenient to introduce complex coordinates
both on the worldsheet and in target space (see
\cite{Novikov:1984ac} for a review of instantons in two--dimensional
sigma models). In the conformal gauge $\sqrt{h}\,
h^{IJ}=\delta^{IJ}$ and in the $(z,\bar z)$ coordinate system on the
worldsheet the action takes the form
\be
S_E=\frac{T}{2}\,\int d^2z\,e_z{}^ie_{\bar z}{}^j\,\delta_{ij}\,.
\ee
To introduce complex coordinates on the target sphere it is
convenient to describe it as $CP^1$. The Fubini-Study metric on $CP^1$ is
\begin{equation}\label{cp1metric}
ds^2_{CP^1}=\frac{d\zeta\,d\bar\zeta}{(1+|\zeta|^2)^2}\,.
\end{equation}
If we choose $\zeta$ to be
\begin{equation}
\zeta=\tan\frac{\theta}{2}\,e^{i\varphi}\,,
\end{equation}
eq. \eqref{cp1metric} takes the form of the metric on $S^2$ of radius
$\frac{1}{2}$
\begin{equation}\label{sigma}
ds^2=\frac{1}{4}(d\theta^2+\sin^2\theta\,d\varphi^2)\,.
\end{equation}
In the $\zeta$, $\bar\zeta$ coordinate system the string action
takes the following form (which is similar to that of the
$O(3)$--sigma model)
\be
S_E=\frac{TR^2}{4}\,\int d^2z\,\frac{|\partial\zeta|^2+|\bar\partial\zeta|^2}{(1+|\zeta|^2)^2}\,.
\ee
It is now obvious that a local minimum is attained if $\bar\partial\zeta=0$
or $\partial\zeta=0$,
\emph{i.e.} the embedding is given by a holomorphic function
$\zeta=\zeta(z)$ for the instanton or by an anti--holomorphic
function $\zeta=\zeta(\bar z)$ for the anti--instanton. The
remaining part of the action can be shown to be a topological
invariant, namely,
\be\label{instac}
S_I=\pi\,n\,{TR^2}=n\,\frac{R^2_{CP^3}}{2\alpha'},
\ee
where $n$ is the topolgical charge of the instanton and
$R_{CP^3}=e^{\frac{1}{3}\phi_0}\,R$ is the $CP^3$ radius in the
string frame.

What we have just reproduced is the classical instanton solution of
the two--dimensional $O(3)$ sigma--model
\cite{Polyakov:1975yp} or rather its extension to $CP^3$
\cite{Golo:1978dd,Golo:1978de,D'Adda:1978uc} which in terms of the Fubini--Studi
coordinates $\zeta^a$ $(a=1,2,3)$ of $CP^3$ (see eq. \eqref{B1} of
Appendix B) has the form
$$
\zeta^a=\zeta^a(z)\quad {\rm or}\quad \zeta^a=\zeta^a(\bar z)\,.
$$
The difference with the $CP^N$ models is that in our case the string
action is also invariant under worldsheet reparametrization. This
means that every classical string solution must satisfy the Virasoro
constraints implying that the worldsheet bosonic physical fields are
associated with the string oscillations transverse to the
worldsheet. For the instanton solution the string excitations along
$AdS_4$ are zero and the Virasoro constraints have the following
form in the conformal gauge
\be\label{virasoro}
\left(\delta^{ab}(1+|\zeta|^2)-{\zeta^b\,\bar\zeta^a}\right)\,\frac{{\partial\zeta^a}\,
\partial\bar\zeta^b}{(1+|\zeta|^2)^2}=0\,.
\ee
We see that the Virasoro constraints are identically satisfied by
the (anti)instanton solution.

Let us note\footnote{We are thankful to Massimo Bianchi for bringing
our attention to this fact.} that though in the $AdS_4
\times CP^3$ background the purely bosonic components of the NS--NS
3--form field strength $H_3$ are zero, the NS--NS 2--form may have
non--zero expectation values  proportional to the K\"ahler two--form
on $CP^3$, $B_{2}=\frac{\alpha'}{R^2}\,a\,J_{a'b'}\,e^{a'}\,e^{b'}$,
where $a$ plays the role of a constant axion.  For such a two--form,
$H_3=dB_2$ is zero since $J_2$ is the closed (but not exact) form,
$dJ_2=0$. In this case also the Wess--Zumino part of the (Wick
rotated) string action
\eqref{cordaA} will contribute to the instanton action, which becomes
\be\label{instac+a}
S_I=n\,(\pi\,R^2\,T-ia)=n\,(\frac{R^2_{CP^3}}{2\alpha'}-ia).
\ee
A similar situation one has in the case of string instantons on
Calabi--Yau spaces \cite{Wen:1985jz,Dine:1986zy}. In the context of
the $AdS_4/CFT_3$ correspondence, the co--homologically non--trivial
$B_2$ field appears from the string side in the generalization of
the ABJM model to include gauge groups of a different rank proposed
in \cite{Aharony:2008gk} (see the Summary below for more discussion
of this point).

Finally, we note that the bosonic string instanton has twelve zero
modes. Four of them correspond to the directions along $AdS_4$ and
eight are the instanton zero modes on $CP^3$
\cite{Perelomov:1987va}. We are now in a position to proceed with
the study of the fermionic zero modes carried by the string
instanton. We shall see that their number is also twelve.

\section{Fermionic equations of motion and the fermionic
zero modes of the string instanton on $CP^3$}\label{fmodes} In a
general supergravity background the equation of motion for the
fermions following from the Green--Schwarz action
\eqref{cordaA} (with the choice of superspace constraints given in
Appendix A.4) is
\begin{eqnarray}\label{fermi1}
0&=&
-\sqrt{-h}h^{IJ}\mathcal E_I{}^\mathcal A T_{\mathcal A\underline\alpha}{}^A\mathcal E_{JA}
-\frac{1}{2}\ve^{IJ}\mathcal E_J{}^{\mathcal B}\mathcal E_I{}^{\mathcal A}H_{\mathcal{AB}\underline\alpha}
\nonumber\\
&&\\
&=&
2i\sqrt{-h}h^{IJ}\mathcal E_J{}^A(\Gamma_A\mathcal E_I)_{\underline\alpha}
+2i\ve^{IJ}\mathcal E_I{}^A(\Gamma_A\Gamma_{11}\mathcal E_J)_{\underline\alpha}
+i\sqrt{-h}h^{IJ}G_{IJ}\lambda_{\underline\alpha}
-i\ve^{IJ}\mathcal E_I{}^A\mathcal E_J{}^B(\Gamma_{AB}\Gamma_{11}\lambda)_{\underline\alpha}\,.\nonumber
\end{eqnarray}
Taking into account that on the mass shell the auxiliary metric
$h_{IJ}$ and the induced metric $G_{IJ}={\mathcal
E}_I{}^A\,{\mathcal E}_J{}^B\,\eta_{AB}$ are proportional to each
other
$$
G_{IJ}=\frac{1}{2}\,h_{IJ}\,(h^{KL}\,G_{KL})\,\quad \Rightarrow
\quad \sqrt{-h}\,h^{IJ} = \sqrt{-G}\,G^{IJ}\quad \Rightarrow \quad \sqrt{-h}\,h^{IJ}\,G_{IJ}=2\sqrt{-G}
$$
the fermionic equations of motion take the following form which
reflects the kappa--symmetry of the theory
\be\label{fermi2}
(1-\Gamma)\,[G^{IJ}\,\mathcal E_J{}^A\,\Gamma_A\mathcal E_I+\lambda]=0\,,
\ee
where $\Gamma$ is the matrix which appears in the kappa--symmetry
projector \eqref{kappastring} and $\lambda^{\underline\alpha}$ is
the dilatino superfield.

For completeness, let us also present the equations of motion of the
string bosonic modes
\be\label{bosonice}
\nabla_I\,(\sqrt{-G}\,G^{IJ}{\mathcal E}_{JA})
+\sqrt{-G}\,G^{IJ}{\mathcal E}_J{}^{\mathcal B}\,T_{\mathcal BA}{}^D
\,{\mathcal E}_{ID}\, +\frac{1}{2}\varepsilon^{IJ}\,{\mathcal
E}_J{}^{\mathcal B}\,{\mathcal E}_I{}^{\mathcal C}\,H_{\mathcal
C\mathcal BA}\,=0,
\ee
where $T_{\mathcal BA}{}^D$ are torsion components and $H_{\mathcal
C\mathcal BA}$ are components of the NS--NS superfield strength with
vector indices $A,D$ and with the indices $\mathcal C$ and
${\mathcal B}$ standing for both the vector and the spinor indices
(see Appendix A.4).

At the linearized level in the $AdS_4\times CP^3$ superspace the
equation of motion for the fermions \eqref{fermi2} reduces to
\begin{eqnarray}\label{fermilinear}
0&=&(1-\Gamma)\big(g^{IJ}e_I{}^A\Gamma_A\mathcal D_J\Theta+\frac{2i}{R}\gamma^5\ups\big)\\
&=& (1-\Gamma)\big(
g^{IJ}e_I{}^A\Gamma_A\nabla_J\Theta+\frac{i}{R}g^{IJ}e_I{}^Ae_J{}^B\Gamma_A\mathcal
P_6\gamma^5\Gamma_B\Theta\big)\,,
\nonumber
\end{eqnarray}
where $g^{IJ}$ is inverse of
\be\label{g}
g_{IJ}=e_I{}^A\,e_J{}^B\,\eta_{AB}=e^{-\frac{2}{3}\phi_0}\,G_{IJ}|_{\Theta=0}
\ee
and
\be\label{nabla}
\nabla_J\Theta=
\left\{\begin{array}{l}
\nabla_J\ups=(\partial_J-\frac{1}{4}\omega_J{}^{ab}\gamma_{ab}-2iA_J\gamma_7)\ups\\
\nabla_J\vartheta=(\partial_J-\frac{1}{4}\omega_J{}^{ab}\gamma_{ab}-\frac{1}{4}\omega_J{}^{a'b'}\gamma_{a'b'})\vartheta\\
\end{array}\right.\,,
\ee
where remember that $A_J=\frac{1}{8}\,\omega_J{}^{a'b'}\,J_{a'b'}$.

Note that one can alternatively derive eq.
\eqref{fermilinear} by varying the quadratic action \eqref{action2}
or its Wick rotated counterpart \eqref{wickra}.

\subsection{Restriction to the instanton solution}\label{fi}
As we have discussed in Section \ref{bi}, the instanton solution is
supported on the $CP^3$ two--dimensional subspace whose tangent
space is characterized
\emph{e.g.} by the first two values of the $CP^3$ tangent space
index $a'=1,2,\ldots,6$. Restricting to this solution we have
\begin{equation}
e_I{}^a=0\,,\qquad e_I{}^{a'}=(e_I{}^i,e_I{}^{\tilde a}=0)\,,\qquad
J_{ij}=\ve_{ij}\,,\qquad i=1,2\quad\mbox{and}\quad\tilde
a=3,4,5,6\,.
\end{equation}
It will be convenient to choose the $CP^3$ gamma matrices as follows
\begin{equation}
\label{gammasigma}
\gamma^{a'}=(\rho^i\otimes\mathbf1,\rho^3\otimes\gamma^{\tilde
a})\,,
\qquad\gamma_7=-\rho^3\otimes\gamma_{\tilde 5}\,,\qquad\gamma_{\tilde 5}
=\frac{1}{4!}\ve_{\tilde a\tilde b\tilde c\tilde d}\gamma^{\tilde
a\tilde b\tilde c\tilde d}\,,
\end{equation}
where $(\rho^1,\rho^2,\rho^3)=(\sigma^1,\sigma^3,-\sigma^2=i\ve)$
are the (re--labeled) Pauli matrices so that $\rho^1\rho^2=i\rho^3$,
and $\gamma^{\tilde a}$ are $4\times4$ Dirac gamma matrices
corresponding to the four--dimensional subspace of $CP^3$ orthogonal
to the instanton $CP^1$ and $\gamma_{\tilde 5}^2=1$.

The Wick rotated kappa--symmetry projection matrix \eqref{gbse} then
reduces to
\begin{equation}\label{gbsi}
\Gamma=i\frac{e^{\frac{2}{3}\phi_0}}{2\sqrt{G}}\,\varepsilon^{IJ}e_I{}^ie_J{}^j\rho_{ij}\rho^3\gamma_5\gamma_{\tilde 5}
=-\frac{\det{e_I{}^i}}{\sqrt{\det{e_I^ie_J^j\delta_{ij}}}}\gamma_5\gamma_{\tilde
5} =-\gamma_5\gamma_{\tilde 5}\,
\end{equation}
and the fermionic part of the Euclidean action \eqref{wickra}
becomes
\be\label{fermiac}
S_F={T}\,\int d^2\xi\,\sqrt
{g}\,g^{IJ}\,\Theta(1-\Gamma)\gamma_5\,\big[i\,e_I{}^i\rho_i\nabla_J\Theta
-\frac{1}{R}e_I{}^ie_J{}^j\rho_i\mathcal P_6\rho_j\Theta\big]\,,
\ee
where the metric $g_{IJ}$ was defined in \eqref{g}. Note that in our
case the fermionic terms of this two--dimensional theory differ from
those of the conventional $2d$ supersymmetric $O(3)\sim CP^1$ (or in
general $CP^N$) sigma--model (see
\cite{Novikov:1984ac, Perelomov:1989im} for a review
and references). For comparison, the $CP^N$ sigma--model Lagrangian
is
\be\label{CPN}
L_{CP^N}= G_{a\bar
b}(\zeta,\bar\zeta)\,\left(\partial_I\bar\zeta^{\bar
b}\,\partial_I\zeta^a+i\Psi^{\dagger\bar
b}\rho^ID_I\Psi^a\right)-\frac{1}{2}\,R_{a\bar bc\bar
d}(\Psi^{\dagger b}\Psi^a)\,(\Psi^{\dagger d}\Psi^c)\,,
\ee
where now $\zeta^a(\xi)$ ($a,\bar a=1,\ldots,N$) are the complex
 $CP^N$ coordinates and $\Psi^a$ and $\Psi^{\dagger
\bar a}$ are independent complex $2N$--component spinor fields,
$G_{a\bar b}(\zeta,\bar\zeta)$ is the K\"ahler (Fubini-Study) metric
on $CP^N$ (see eq.
\eqref{B1} of Appendix B for the $CP^3$ case),
$D_I\,\Psi^a=\partial_I\,\Psi^a+\Gamma^a_{bc}\,\partial_I\,\zeta^b\,\Psi^c$
and $\Gamma^a_{bc}$ and $R_{a\bar bc\bar d}$ are the $CP^N$
Christoffel symbol and curvature, respectively.

In view of the form of the quadratic action \eqref{fermiac} and of
the fermionic equation \eqref{fermilinear} it is natural to impose
on the fermionic fields the conventional kappa--symmetry
gauge--fixing condition
\begin{equation}\label{kappafixing}
\frac{1}{2}(1+\Gamma)\Theta=\frac{1}{2}(1-\gamma_5\gamma_{\tilde 5})\Theta=0\,,
\end{equation}
which means that the fermions split into two sectors according to
their chiralities in $AdS_4$ and in the four--dimensional subspace
of $CP^3$ orthogonal to the instanton $CP^1$
\begin{equation}
\Theta_+:\quad\gamma_5\Theta_+=\gamma_{\tilde 5}\Theta_+=\Theta_+\,,\qquad
\Theta_-:\quad\gamma_5\Theta_-=\gamma_{\tilde 5}\Theta_-=-\Theta_-\,.
\end{equation}
Using the form of the $CP^3$ gamma--matrices \eqref{gammasigma} we
find that
\begin{equation}
J=-iJ_{a'b'}\gamma^{a'b'}\gamma_7 =-2\gamma_{\tilde 5}+iJ_{\tilde
a\tilde b}\gamma^{\tilde a\tilde b}\gamma_{\tilde 5}\rho^3
=-2\gamma_{\tilde 5}+2\rho^3\tilde J(1-\gamma_{\tilde 5})\,,
\end{equation}
where
\begin{equation}
\tilde J=-\frac{i}{4}J_{\tilde a\tilde b}\gamma^{\tilde a\tilde b}=-\frac{i}{8}J_{\tilde a\tilde b}\gamma^{\tilde a\tilde b}(1-\gamma_{\tilde 5})
\qquad\tilde J^2=\frac{1}{2}(1-\gamma_{\tilde 5})\,.
\end{equation}
So, the supersymmetry projection matrices $\mathcal P_2$ and
$\mathcal P_6$ become
\begin{eqnarray}\label{p2p6}
\mathcal P_2&=&\frac{1}{8}(2+J)=\frac{1}{4}(1+\rho^3\tilde J)(1-\gamma_{\tilde 5})\nonumber\\
\mathcal P_6&=&\frac{1}{8}(6-J)=\frac{1}{4}\big(3+\gamma_{\tilde 5}-\rho^3\tilde J(1-\gamma_{\tilde 5})\big)\,.
\end{eqnarray}
Their action on the two sets of the chiral fermions is
\be\label{projaction1}
\mathcal P_6\Theta_+=\Theta_+=\vartheta_+\,,\qquad \mathcal
P_2\Theta_+=\upsilon_+=0\,,
\ee
\be\label{projaction2}
 \mathcal P_6\Theta_-=\frac{1}{2}(1-\rho^3\tilde J)\Theta_-=\vartheta_-\,,\qquad
 \mathcal P_2\Theta_-=\frac{1}{2}(1+\rho^3\tilde J)\Theta_-=\ups\,.
\ee
Note that from eqs. \eqref{projaction1} and \eqref{projaction2} it
follows that all the eight $\vartheta_+$ are fermions which
correspond to unbroken supersymmetries of the $AdS_4\times CP^3$
superbackground, while in the $\Theta_-$ sector four fermions
($\vartheta_-$) correspond to unbroken supersymmetries and other
four ($\ups$) to the broken ones. Note also that since for the
instanton configuration the kappa--symmetry projector \eqref{gbsi}
commutes with the `supersymmetry' projectors \eqref{p2p6}, it is not
possible to choose the kappa--symmetry gauge--fixing condition which
would put to zero all the eight `broken--supersymmetry' fermions. In
terms of the fields $\vartheta_+$, $\vartheta_-$ and $\upsilon$ the
fermionic action \eqref{fermiac} takes the form
\be\label{fermiac1}
S_F={2T}\int d^2\xi\,\det
{e}\,\left[i\,\vartheta_+e_i{}^I\,\rho^i\nabla_I\vartheta_+
-\frac{2}{R}\vartheta_+\vartheta_+
-2\big(i\,\upsilon\,e_i{}^I\,\rho^i\nabla_I\vartheta_-
-\frac{1}{R}\upsilon\upsilon\big)\right]\,,
\ee
where $e_i{}^I$ is the inverse vielbein on $S^2$.

For the instanton configuration the fermionic equation
\eqref{fermilinear} reduces to the following ones
\bee
\label{Dirac1}
e_i{}^I\,\rho^i\nabla_I\vartheta_++\frac{2i}{R}\vartheta_+
&=&0\,,\\
\label{Dirac2}
e_i{}^I\,\rho^i\nabla_I\vartheta_-+\frac{2i}{R}\ups&=&0\,,\\
\label{Dirac3}
e_i{}^I\,\rho^i\nabla_I\ups&=&0\,.
\eee
{}From the form of the action \eqref{fermiac1} and the equation of
motion \eqref{Dirac2} it follows that the field $\upsilon$ can be
regarded as an auxiliary one, which can be expressed in terms of a
derivative of $\vartheta_-$. However, for the analysis of the
solutions of eqs.
\eqref{Dirac1}--\eqref{Dirac3} it is more convenient to consider it
as an independent variable satisfying the Dirac equation
\eqref{Dirac3}.

The covariant derivative $\nabla_I$ (defined in
\eqref{nabla}) contains the pullback on the instanton two--sphere of the $CP^3$
spin connection whose explicit form is given in Appendix B
\be\label{nablai}
\nabla_I\vartheta_{\pm}=(\partial_I-\frac{1}{4}\,\omega_I{}^{a'b'}\,\gamma_{a'b'})\,\vartheta_{\pm}\,.
\ee
Computing the pullback of the $CP^3$ connection, substituting it
into the Dirac equations and taking into account the projection
properties of the spinors we get the fermionic equations in the
following form
\bee
\label{Dirac11}
e_i{}^I\rho^i\,\nabla^{S^2}_I\vartheta_++\frac{2i}{R}\vartheta_+=
e_i{}^I\rho^i(\nabla^{S^2}_I\vartheta_++\frac{i}{R}\,e_I{}^j\rho_j\vartheta_+)
&=&0\,,\\
\label{Dirac21}
e_i{}^I\rho^i(\nabla^{S^2}_I+i\tilde A_I\,\rho^3)\vartheta_-+\frac{2i}{R}\ups&=&0\,,\\
\label{Dirac31}
e_i{}^I\rho^i\,(\nabla^{S^2}_I-i\tilde A_I\,\rho^3)\,\ups&=&0\,,
\eee
where $\nabla_{S^2}=d-\frac{1}{4}\omega^{ij}_{S^2}\,\rho_{ij}$ is
the intrinsic covariant derivative on the sphere of radius
$R_{S^2}=R/2$ with curvature
$R^{ij}_{S^2}=d\omega^{ij}_{S^2}=\frac{4}{R^2} e^i\,e^j$ and $\tilde
A$ can be interpreted as the electromagnetic potential induced by a
magnetic monopole of charge $g=-1/2$ placed at the center of the
sphere. This is due to the fact that
\begin{equation}
F=d\tilde
A=\frac{1}{R^2}\,e^ie^j\ve_{ij}=\frac{1}{2}\,e^ie^j\,F_{ji}\quad\Rightarrow\quad
F_{ij}=-\frac{2}{R^2}\,\ve_{ij}=\frac{g}{R_{S^2}^2}\ve_{ij}\,.
\end{equation}
Note that $\frac{1}{4}\,\omega^{ij}_{S^2}\varepsilon_{ij}$ and
$\tilde A$ are equivalent up to a total derivative term
$$
\tilde A=\frac{1}{4}\,\omega^{ij}_{S^2}\varepsilon_{ij}+d\Lambda\,.
$$
In our parametrization of $CP^3$ (see Appendix B) and for a given
embedding of $S^2$ in $CP^3$, $\omega_{S^2}$ and $\tilde A$ have the
following form in terms of the angular coordinates on $S^2$
\be\label{oA}
\omega^{12}_{S^2}=\cos\theta d\varphi\,,\qquad \tilde A=\frac{1}{2}(1+\cos\theta)\,d\varphi\,.
\ee
We are now in a position to analyze the solutions of the fermionic
equations \eqref{Dirac11}--\eqref{Dirac31}. Eq.
\eqref{Dirac11} has the form of the Dirac equation for a fermion of mass $\frac{2}{R}$. It is the product of $e_i{}^I\rho^i$ with the Killing
spinor equation on the sphere
\be\label{Killing}
(\nabla^{S^2}_I+\frac{i}{R}\,e_I{}^j\rho_j)\vartheta_+=0\,.
\ee
The Killing spinor equation on $S^2$ for a two--component spinor has
two non--trivial solutions \cite{Fujii:1985bg}. Our $\vartheta_+$
spinors carry four (independent) external indices in addition to the
$S^2$--spinor index. Therefore, eq. \eqref{Killing} has eight
solutions which are obviously solutions of the Dirac equation
\eqref{Dirac11}. These are actually the only regular eigenspinors of
the Dirac operator on the sphere with the eigenvalue $-2i/R$
\cite{Abrikosov:2002jr}. Thus, in the $\Theta_+$ sector the string
instanton has eight fermionic zero modes which are the solutions of
the Killing spinor equation
\eqref{Killing}. In spherical coordinates they have the explicit
form
\cite{Lu:1998nu}
\begin{equation}
\vartheta_+=
e^{-\frac{i}{2}\theta\rho^1}e^{\frac{i}{2}\varphi\rho^3}\epsilon_+
=\left(\cos{\frac{\theta}{2}}-i\rho^1\sin{\frac{\theta}{2}}\right)\left(\cos{\frac{\varphi}{2}}+i\rho^3\sin{\frac{\varphi}{2}}\right)\epsilon_+\,,
\end{equation}
where $\epsilon_+$ is an arbitrary constant spinor satisfying the
chirality conditions $\gamma_5\epsilon_+=\gamma_{\tilde
5}\epsilon_+=\epsilon_+$.

Let us now proceed with the analysis of the third fermionic equation
\eqref{Dirac31}. As we have already mentioned, this equation describes
the electric coupling of the fermionic field $\ups$ to the monopole
potential on the sphere. The electric charge of $\ups$ is  $e=\pm 1$
for $\ups=\pm\rho_3\ups$, \emph{i.e.} when $\ups$ is a
chiral/anti--chiral two--dimensional spinor, respectively.  The
analysis in
\cite{Deguchi:2005qd} then tells us that there are non--trivial
solutions of the charged Dirac equation \eqref{Dirac31} of positive
chirality when $ge\geq1/2$ and of negative chirality when
$ge\leq-1/2$. Since we are in the opposite situation, there are no
non--trivial solutions in our case and hence $\ups=0$.

If $\ups=0$, eq. \eqref{Dirac21} implies that $\vartheta_-$ should
satisfy the massless Dirac equation
\be\label{Dirac22}
e_i{}^I\rho^i(\nabla^{S^2}_I+i\tilde
A_I\,\rho^3)\vartheta_-=e_i{}^I\rho^i(\partial_I
+\frac{i}{2}\rho_3\partial_I\varphi)\vartheta_-=0\,.
\ee
We observe that the electric charge of $\vartheta_-$ is opposite to
that of $\upsilon$, \emph{i.e.} it is $e=\mp 1$ depending on whether
$\vartheta_-$ is chiral or anti--chiral two--dimensional spinor,
\emph{i.e.} whether $\vartheta_-=\pm\rho_3\vartheta_-$. Now we are in the
situation in which the requirement of \cite{Deguchi:2005qd} for the
Dirac equation \eqref{Dirac22} to have non--trivial solutions is
saturated, \emph{i.e.} in our case for $\vartheta_-$ of positive
$\rho^3$--chirality $ge=1/2$ and for $\vartheta_-$ of negative
$\rho^3$--chirality $ge=-1/2$. By the Atiyah--Singer index theorem
there is one solution for each $\rho^3$--chirality of $\vartheta_-$.
The general solution of \eqref{Dirac22} has actually a very simple
form
\be\label{sol-}
\vartheta_-=\frac{1}{2}\,e^{-\frac{i}{2}\rho_3\,\varphi}\,\big[(1+\rho^3)\lambda_-(\zeta)
+(1-\rho^3)\,\mu_-(\bar\zeta)\big]\,,
\ee
where $\lambda_-(\zeta)$ and $\mu_-(\bar\zeta)$ are holomorphic and
anti--holomorphic spinors in the projective coordinates $\zeta$ and
$\bar\zeta$ of $S^2\simeq CP^1$ which are anti--chiral in the
directions transverse to the instanton, \emph{i.e.}
$\lambda_-=-\gamma_5\lambda_-$, $\lambda_-=-\gamma_{\tilde
5}\lambda_-$\, $\mu_-=-\gamma_5\mu_-$ and $\mu_-=-\gamma_{\tilde
5}\mu_-$. For the anti--instanton the solution takes the same form
but with anti--holomorphic $\lambda_-(\bar\zeta)$ and holomorphic
$\mu_-(\zeta)$.

In \cite{Deguchi:2005qd} it has been shown that the only
normalizable solutions of the Dirac equation \eqref{Dirac22} are
those with constant $\lambda_-$ and $\mu_-$ in \eqref{sol-}. This
allows us to conclude that in the $\vartheta_-$ sector the string
instanton has four zero modes characterized by eq. \eqref{sol-} with
constant $\lambda_-$ and $\mu_-$.\footnote{Remember that the
eight--component spinor $\vartheta_-$ satisfies the additional
projection condition
\eqref{projaction2} which reduces the number of its components to
four.} Note that for $\lambda_-=const$ and $\mu_-=const$ the spinor
\eqref{sol-} is the solution of the stronger equation
\be\label{stronger}
(\partial_I +\frac{i}{2}\rho_3\partial_I\varphi)\vartheta_-=0\,.
\ee
This equation is the projection on the instantonic sphere of the
$AdS_4\times CP^3$ Killing spinor equation for $\vartheta_-$.

 To summarize, when $\upsilon=0$ and in view of the form of
the fermionic supervielbeins $E^{\alpha a'}$  $(a'=1,\ldots, 6)$
\footnote{To avoid confusion, let us note that the index $a'$ on
spinors is different from the same index on bosonic quantities. See
the end of Appendix A.5 for a more detailed explanation.} of the
supercoset $OSp(6|4)/U(3)\times SO(1,3)$ (see Appendix A.7), the
non--linear fermionic equation of motion
\eqref{fermi2} as well as the linear one
\eqref{fermilinear} involve the pull--back on the string worldsheet of
the $AdS_4\times CP^3$ Killing spinor operator
\be\label{kllingo}
{\mathcal D}\vartheta = D_{24}\vartheta={\mathcal P_6}\,(d
+\frac{i}{R}\,e^{ a}\gamma^5\gamma_{  a}
+\frac{i}{R}e^{a'}\gamma_{a'}-\frac{1}{4}\omega^{  a b}\gamma_{ a b}
-\frac{1}{4}\omega^{a'b'}\gamma_{a'b'})\vartheta\,,
\ee
which acts on the 24 fermions $\vartheta$ associated with the
supersymmetry of $AdS_4\times CP^3$ (see Appendix A.7). Therefore,
if $\vartheta$ are the 24 Killing spinors on $AdS_4\times CP^3$ they
solve not only the linearized equations \eqref{fermilinear} but also
the complete fermionic equations \eqref{fermi2}. In the case of the
string instanton considered above, the kappa--symmetry projector
reduces the number of solutions of the pulled--back Killing spinor
equation by half, leaving us with the twelve physical fermionic zero
modes. It should also be noted that these fermionic zero modes do
not contribute to the bosonic equations
\eqref{bosonice}. This guarantees that the bosonic instanton
solution does not have a back reaction from the fermionic modes.

We should note that the Dirac equations
\eqref{Dirac11}--\eqref{Dirac31} may have (non--normalizable)
solutions which are not the Killing spinors (as \emph{e.g.} eq.
\eqref{sol-} with non--constant $\lambda$ and $\mu$).
However, these other fermionic modes would modify the string field
equations at higher order in fermions. In particular, they would
produce a non--trivial contribution to the bosonic field equations
\eqref{bosonice},
\emph{i.e.} back--react on the form of the purely bosonic instanton
and, hence, should be discarded.

Let us stress once again that, as we have shown, for the instanton
solution considered above the kappa--symmetry cannot eliminate all
the eight fermions $\upsilon$ associated with the supersymmetries
broken in the $AdS_4\times CP^3$ background. Therefore, even if
among the instanton fermionic zero modes there is no
$\upsilon$--modes, the fluctuations around the instanton solution
will have four physical fermionic degrees of freedom corresponding
to the target--space supersymmetries broken by the $AdS_4\times
CP^3$ background.

\subsection{Fermionic zero modes and supersymmetry}
Let us discuss in more detail how the fermionic zero modes are
related to supersymmetry of the $AdS_4\times CP^3$ superbackground
and, correspondingly, of the superstring action. At the linearized
level in fermions the supersymmetry part of the $OSp(6|4)$
transformations acts as follows
\begin{eqnarray}\label{susy1}
\delta\vartheta&=&\epsilon\,,\nonumber\\
\delta\upsilon&=&0\,,\nonumber\\
\delta X^M e_M{}^A(X)&=&-i\epsilon\Gamma^A\vartheta\,,
\end{eqnarray}
where $\epsilon\equiv\mathcal P_6\,\epsilon(X)$ are 24 supersymmetry
parameters of $OSp(6|4)$ satisfying the $AdS_4\times CP^3$ Killing
spinor equation
\begin{equation}
\mathcal D\epsilon=\nabla\epsilon+\frac{i}{R}e^A\,\mathcal
P_6\gamma^5\Gamma_A\epsilon=0
\end{equation}
with the explicit form of $\mathcal D$ given in eq. \eqref{kllingo}.
Note that, at the leading order in fermions, the eight fermions
$\upsilon$ are not subject to the supersymmetry transformations. The
action of the isometry group $OSp(6|4)$ on these fermions is such
that it takes the form of induced $SO(1,3)\times U(1)$ rotations
with parameters depending on $X$, $\vartheta$ and the $OSp(6|4)$
parameters
\be\label{deltav}
\delta \upsilon
=\frac{1}{4}\,\Lambda_{AB}(\epsilon,X,\vartheta)\,\Gamma^{AB}\,\upsilon\,.
\ee
 Thus the first nontrivial term in the supersymmetry
variation of $\upsilon$ is quadratic in fermionic fields which is
beyond the linear approximation we are interested in.

It is not hard to see that the quadratic string action
\eqref{action2} is invariant under the supersymmetry transformations
\eqref{susy1} (up to quadratic order in fermions). At the same time,
the action \eqref{wickra}, which is obtained from \eqref{action2} by
the shift \eqref{y} of the $CP^3$ coordinates, is invariant under
the supersymmetry with the transformations of the\emph{ shifted}
bosonic coordinates being
\be\label{xhat}
\delta \hat{X}^M e_M{}^A(\hat
X)=-i\epsilon\Gamma^A\vartheta-i\epsilon\Gamma^A\upsilon\,=-i\epsilon\Gamma^A\Theta\,.
\ee
Let us now briefly recall how the target--space supersymmetry gets
converted into worldsheet supersymmetry upon elimination of the
un--physical fermionic degrees of freedom by gauge fixing
kappa--symmetry. A more detailed discussion of such a
``transmutation" of supersymmetry and its partial breaking in the
Green--Schwarz formulation of superstrings and superbranes the
reader may find\emph{ e.g.} in
\cite{Hughes:1986dn,Hughes:1986fa,Bergshoeff:1997kr,Claus:1998yw}.

If we impose on the fermionic fields $\Theta=(\vartheta,\,\upsilon)$
a kappa--symmetry gauge condition as \emph{e.g.} the one we have
used for studying the instanton solution, eq. \eqref{kappafixing},
\be\label{kappafixing1}
\frac{1}{2}(1+\Gamma_0)\,\Theta=0\,,
\ee
the kappa--symmetry gauge--fixing condition will not be invariant
under all the twenty--four supersymmetries \eqref{susy1} but only
under half of them satisfying the condition
\be\label{unbroken}
\epsilon_{br}=\frac{1}{2}(1-\Gamma_0)\,\epsilon\,.
\ee
In eqs. \eqref{kappafixing1} and \eqref{unbroken} we denoted the
gauge--fixing projector by $\Gamma_0$ to distinguish it from the
more general projector matrix $\Gamma$ that appears in the
kappa--symmetry transformations \eqref{kappastring}--\eqref{gbs}.

The target--space supersymmetries with the parameter $\epsilon_{br}$
are those which are spontaneously broken by the presence of the
string. The reason is that the remaining twelve fermionic fields
$\vartheta=\frac{1}{2}(1-\Gamma_0)\,\vartheta$ get shifted by these
transformations and hence behave as Volkov--Akulov goldstinos
\cite{Volkov:1972jx,Volkov:1973ix}.

The supersymmetries which remain unbroken and which become
worldsheet supersymmetries are identified as follows. The gauge
fixing condition \eqref{kappafixing1} is not invariant under the
supersymmetry transformations \eqref{susy1} with the parameter
$\epsilon_w=\frac{1}{2}(1+\Gamma_0)\,\epsilon$. However, this can be
cured by an appropriate compensating kappa--symmetry transformation
that (at the leading order in fermions) satisfies the condition
\be\label{compens}
-\frac{1}{4}\,{\mathcal P}_6(1+\Gamma_0)(1+\Gamma)\,\kappa =
\epsilon_w\equiv
\frac{1}{2}(1+\Gamma_0)\,\epsilon.
\ee
This condition relates the components of the $\kappa$--symmetry
parameter appearing in the transformation of $\vartheta$ to the
supersymmetry parameter $\epsilon_w$. Since kappa--symmetry is the
worldsheet fermionic symmetry which can actually be identified with
the conventional local worldsheet supersymmetry
\cite{Sorokin:1999jx}, eq. \eqref{compens} thus converts the unbroken target--space
supersymmetries into worldsheet supersymmetry. Note that eq.
\eqref{compens} does not involve the part of the kappa--symmetry
transformation acting on the $\upsilon$--fermions since they are
singled out with the complementary projector $\mathcal P_2$. This
part of kappa--symmetry is fixed by the gauge condition
$\frac{1}{2}(1+\Gamma_0)\upsilon=0$ (see eq.
\eqref{kappafixing1}).

As a result, (at a leading order in fermions and bosons) under the
broken and unbroken supersymmetries the worldsheet fermionic
fields remaining after the gauge-fixing \eqref{kappafixing1} $\Theta\equiv\frac{1}{2}\,(1-\Gamma_0)\Theta=(\vartheta,\upsilon)$
and the bosonic fields $\hat X^M$  transform as follows
\bee
\delta\upsilon&=&0\,,\qquad
\delta\vartheta=\epsilon_{br}+\frac{1}{4}\,{\mathcal P}_6\,(1-\Gamma_0)(1+\Gamma)\kappa\,,\label{xhat10}\\
\label{xhat11}
\delta \hat{X}^M\,e_M{}^{i}(\hat X) &=&-i\epsilon_{br}\Gamma^i\Theta-\delta_\kappa\vartheta^{\underline\alpha}\,
E_{\underline
\alpha}{}^{i}(\hat X,\Theta)+\mathcal O(\epsilon,\Theta,\hat X)\,,\\
\label{xhat1}
\delta \hat{X}^M\,e_M{}^{\perp}(\hat X) &=&-i\epsilon_w\Gamma^\perp\Theta-\delta_\kappa\vartheta^{\underline\alpha}\,
E_{\underline
\alpha}{}^{\perp}(\hat X,\Theta)+\mathcal O(\epsilon,\Theta,\hat
X)\,,
\eee
where $i=0,1$ and $\perp=2,\ldots, 9$  indicate the directions
parallel and orthogonal to the string worldsheet, respectively, the
second terms in
\eqref{xhat11} and \eqref{xhat1} come from the compensating kappa--symmetry
transformation
\eqref{kappastring} that at the linearized level is just
$-i\epsilon\Gamma^A\Theta$, and $O(\epsilon,\Theta,\hat X)$ stand
for terms which are non--linear in fields (and their derivatives).

It is instructive to notice that the leading (linear) term in the
supersymmetry transformations of $\hat X^M$ along the directions
trasverse to the string worldsheet contains the parameter of the
unbroken supersymmetries, while along the worldsheet the linear term
contains the broken supersymmetry parameter. This reflects the fact
that the bosonic excitations transversal to the classical string
configuration and kappa--gauge fixed fermionic fields are associated
with worldsheet physical fields forming supermultiplets under the
unbroken supersymmetry. At the same time the supersymmetry
transformations along the string worldsheet can be compensated by an
appropriated worldsheet reparametrization.

For the instanton solution under consideration we have
$\Gamma=\Gamma_0=-\gamma_5\gamma_{\tilde 5}$ and $\upsilon=0$. So
the supersymmetry transformations (at the leading order) become
\bee
\delta\upsilon &=&0+\dots\,,\qquad
\delta\vartheta=\epsilon_{br}+\dots\,,\label{brsusy}\\
\label{xhat2} \delta
\hat{X}^M\,e_M{}^{\perp}(X)&=&-2i\epsilon_w\Gamma^{\perp}\vartheta\,+\,\dots,
\eee
where the dots stand for higher order terms in fermions.

Under the unbroken supersymmetry transformations the fermionic zero
modes induce an (isometry) transformation of the string coordinates
in the transverse directions which results in a shift of the bosonic
parameters characterizing the string instanton. This is analogous to
the supersymmetry transformations of the `collective coordinates' of
the $CP^N$ sigma--model instanton \cite{Novikov:1984ac}.

{}From eqs. \eqref{brsusy} and \eqref{xhat2} we also see that if we
start from the purely bosonic instanton solution discussed in
Section
\ref{bi}, we can find at least part of the instanton fermionic zero
modes by looking at the variation of the fermionic fields under
supersymmetry. The form of the supersymmetry transformations implies
that the bosonic instanton configuration is 1/2 BPS. Namely, the
string instanton solution with $\Theta=0$ is invariant under the
twelve supersymmetries $\epsilon_w$. Fermionic zero modes are
generated by the target--space supersymmetries (with the parameter
$\epsilon_{br}$) which are broken by the string configuration, as we
have already discussed in the end of Section
\ref{fi} where we have also shown that the instanton does not have
other fermionic zero modes associated with the fields $\upsilon$.
Note that the latter cannot be obtained from the purely bosonic
solution by a supersymmetry transformation since the corresponding
variation of $\upsilon$ is proportional to $\upsilon$ itself (see eq.
\eqref{deltav}).

Let us now compare our $AdS_4\times CP^3$ superstring worldsheet
theory (which has $12$ unbroken worldsheet supersymmetries) with the
supersymmetry properties of the conventional $n=(2,2)$\footnote{$n$
labels the real number of left-- and right--handed worldsheet
supersymmetries.} supersymmetric $CP^N$ sigma--model (described by
the Lagrangian
\eqref{CPN}) and with its instanton solutions
\cite{Novikov:1984ac,Perelomov:1989im}.

The supersymmetry transformations in the $CP^N$ sigma--model  have
the following form
\be\label{sigmasusy}
\delta\zeta^a=\bar\epsilon\Psi^a\,, \qquad
\delta\Psi^a=i\rho^I\epsilon\,\partial_I\zeta^a\,+\,\cdots \,, \qquad (a=1,\ldots, N)
\ee
where $\epsilon$ is now a constant complex two--component spinor
parameter of  $n=(2,2)$ supersymmetry and the dots stand for the
terms non--linear in the fields. The $CP^N$ sigma--model is also
invariant under superconformal transformations \cite{D'Adda:1982eh}
\be\label{sigmaconfsusy}
\delta\zeta^a=\bar\eta(z,\bar z)\,\Psi^a\,, \qquad
\delta\Psi^a=i\rho^I\eta(z,\bar z)\,\partial_I\zeta^a\,+\cdots\,, \qquad (a=1,\ldots, N)
\ee
The superconformal transformations are similar to the rigid
supersymmetries \eqref{sigmasusy} but with the complex
two--component spinor parameters whose chiral and anti--chiral
components are, respectively, holomorphic and anti--holomorphic
$\eta(z,\bar z)\,=(\eta_+(z),\eta_-(\bar z))$.

The superconformal symmetry of the $CP^N$ sigma--model (which is
broken by quantum anomalies \cite{Novikov:1984ac}) is in a sense a
counterpart of the spontaneously broken part of the target--space
supersymmetry of the superstring action.

If one starts from the purely bosonic instanton solution of the
$CP^N$ sigma--model
\be\label{cpninst}
\bar\partial\zeta^a=0\,\quad {\rm or} \quad \partial\zeta^a=0\quad
{\rm and} \quad \Psi=0
\ee
 one can then generate solutions of the fermionic field equations
and find the corresponding fermionic zero modes by considering the
supersymmetry transformations
\eqref{sigmasusy} and \eqref{sigmaconfsusy} of $\Psi$. In this way,
taking into account that for the instanton the fields $\zeta^a$ are
either holomorphic or anti--holomorphic, one finds that only half of
the supersymmetry transformations \eqref{sigmasusy} and of the
superconformal transformations \eqref{sigmaconfsusy} are
non--trivial, those with the parameters $\epsilon$ and $\eta$ being
(anti)chiral $2d$ spinors. The fermionic zero modes obtained in this
way are (anti)holomorphic (anti)chiral $2d$ spinor fields. We
observe that in contrast to the case of the string instanton whose
fermionic zero modes are generated by the spontaneously broken
supersymmetry transformations, in the case of the $CP^N$
sigma--model half of the fermionic zero modes are generated by the
rigid supersymmetry transformations and another half by the
superconformal symmetry.

\section{Summary and Discussion}\label{summary}

We have thus found that the string instanton wrapping the
non--trivial two--cycle inside $CP^3$ has twelve fermionic zero
modes. As we have already mentioned, the eight string fermionic
fields $\vartheta_+$ which have an effective mass $\frac{2}{R}$ and
four massless $\vartheta_-$ electrically coupled to the $S^2$
monopole field, correspond to twelve (of the twenty four)
supersymmetries of the $AdS_4\times CP^3$ background. The fermionic
zero modes thus play the role similar to Volkov--Akulov goldstinos
\cite{Volkov:1972jx,Volkov:1973ix} and manifest partial breaking of supersymmetry.
Note that in $AdS_4\times CP^3$ there also exists an NS5--brane
instanton wrapping the entire $CP^3$. It would be of interest to analyze
possible effects of these instantons in $AdS_4\times CP^3$
superstring theory and to understand their counterparts in the
boundary $CFT_3$ theory.

As was mentioned briefly in Section \ref{bipart}  the instanton
solution can be generalized by switching on a NS--NS field $B_2
\sim J_2$ of a non--trivial co--homology on $CP^1\simeq S^2$. The
coupling of the string to the $B$--field then results in the
instanton action being shifted by a constant imaginary piece. In
fact, the co--homologically non--trivial $B$--field arises on the
string side of the $AdS_4/CFT_3$ correspondence when one considers
the ABJ--model \cite{Aharony:2008gk} which generalizes  the ABJM
construction to gauge groups of different rank, \emph{i.e.}
$U(N+l)_k\times U(N)_{-k}$ with $0<l<k$ instead of $U(N)_k\times
U(N)_{-k}$. The appearance of the $B$--field in the string action
thus results in breaking the parity--invariance of the theory. In
\cite{Aharony:2008gk} it has been shown that the integral of $B_2$
on the $CP^1$ cycle inside $CP^3$ takes a fractional value
\be
\frac{1}{2\pi}\int_{CP^1}\,B_2=\frac{l}{k}\,.
\ee
In eleven dimensions this corresponds to a co--homologically
nontrivial three--form potential on the 3--cycle $S^3/\mathbb
Z_k\subset S^7/\mathbb Z_k$. From the point of view of M2--branes
probing a $\mathbb C^4/\mathbb Z_k$ singularity this is associated
to $l$ fractional M2--branes sitting at the singularity. The
fractional M2-branes can be thought of as M5--branes wrapping the
corresponding vanishing 3-cycle at the orbifold point, see
\cite{Aharony:2008gk}.

This picture suggests that the string instanton considered in this
paper should correspond in M--theory to an instantonic M2--brane,
\emph{i.e.} an M2--brane whose worldvolume wraps a 3--cycle
$S^3/\mathbb Z_k\subset S^7/\mathbb Z_k$. There is, however, a
subtlety here. Namely, while in the case of the $D=10$ type IIA
string instanton there are an infinite number
($|n|=1,2,\ldots,\infty$) of topologically different configurations,
the number of non--equivalent M2--brane configurations wrapping the
3--cycle in $S^7/\mathbb Z_k$ of the $D=11$ theory is $k$, since
$H_3(S^7/\mathbb Z_k,\mathbb Z)=\mathbb Z_k$. The reason is that the
string instanton solution has been considered in the $AdS_4\times
CP^3$ background of the pure type IIA $D=10$ supergravity,
\emph{i.e.} in the limit in which (from the $D=11$ perspective) the
radius of the $S^1$ fiber of $S^7$ tends to zero ($k\rightarrow
\infty$). The consideration at finite $k$ would require taking into
account Kaluza--Klein modes and D--brane effects.

It would be interesting to find out what these instantonic strings,
M2--branes and NS5--branes correspond to in the ABJ/ABJM gauge--theory
picture.

\section*{Acknowledgements}
The authors are grateful to Konstantin Zarembo for the suggestion to
look at this stringy instanton problem and for valuable discussions
and comments. They are also thankful to Massimo Bianchi, Marco
Matone, Nikita Nekrasov and Mario Tonin for useful comments and
discussion. This work was partially supported by the INFN Special
Initiative TV12. D.S. was also partially supported by an Excellence
Grant of Fondazione Cariparo (Padova) and the grant FIS2008-1980 of
the Spanish Ministry of Science and Innovation.

\def\thesection{}
\def\theequation{A.\arabic{equation}}\label{A}
\section{Appendix A. Main notation and conventions}
\setcounter{equation}0

The convention for the ten and eleven dimensional metrics is the
`almost plus' signature $(-,+,\cdots,+)$. Generically, the tangent
space vector indices are labeled by letters from the beginning of
the Latin alphabet,  while  letters from the middle of the Latin
alphabet stand for curved (world) indices. The spinor indices are
labeled by Greek letters.

\def\thesubsection{A.1}
\subsection{$AdS_4$ space}

$AdS_4$ is parametrized by the coordinates $x^{  m}$ and its
vielbeins are $e^{  a}=dx^{  m}\,e_{  m}{}^{  a}(x)$, ${
m}=0,1,2,3;$ ${  a}=0,1,2,3$. The $D=4$ gamma--matrices satisfy:
\be\label{gammaa}
\{\gamma^{  a},\gamma^{  b}\}=2\,\eta^{  a  b}\,,
\qquad \eta^{  a  b}={\rm diag}\,(-,+,+,+)\,,
\ee
\be\label{gamma5}
\gamma^5=i\gamma^0\,\gamma^1\,\gamma^2\,\gamma^3, \qquad
\gamma^5\,\gamma^5=1\,.
\ee
The charge conjugation matrix $C$ is antisymmetric, the matrices
$(\gamma^{  a})_{\alpha\beta}\equiv (C\,\gamma^{ a})_{\alpha\beta}$
and $(\gamma^{  a  b})_{\alpha\beta}\equiv(C\,\gamma^{  a
b})_{\alpha\beta}$ are symmetric and $\gamma^5_{\alpha\beta}\equiv
(C\gamma^5)_{\alpha\beta}$ is antisymmetric, with
$\alpha,\beta=1,2,3,4$ being the indices of a 4--dimensional spinor
representation of $SO(1,3)$ or $SO(2,3)$.

\def\thesubsection{A.2}
\subsection{$CP^3$ space}

$CP^3$ is parametrized by the coordinates $y^{m'}$ and its vielbeins
are $e^{a'}=dy^{m'}e_{m'}{}^{a'}(y)$, ${m'}=1,\cdots,6;$
${a'}=1,\cdots,6$. The $D=6$ gamma--matrices satisfy:
\be\label{gammaa'}
\{\gamma^{a'},\gamma^{b'}\}=2\,\delta^{{a'}{b'}}\,,\qquad \delta^{a'b'}={\rm
diag}\,(+,+,+,+,+,+)\,,
\ee
\be\label{gamma7}
\gamma^7=\frac{i}{6!}\,\varepsilon_{\,a_1'a_2'a_3'a_4'a_5'a_6'}\,\gamma^{a_1'}\cdots \gamma^{a_6'} \qquad
\gamma^7\,\gamma^7=1\,.
\ee
The charge conjugation matrix $C'$ is symmetric and the matrices
$(\gamma^{a'})_{\alpha'\beta'}\equiv
(C\,\gamma^{a'})_{\alpha'\beta'}$ and
$(\gamma^{a'b'})_{\alpha'\beta'}\equiv(C'\,\gamma^{a'b'})_{\alpha'\beta'}$
are antisymmetric, with $\alpha',\beta'=1,\cdots,8$ being the
indices of an 8--dimensional spinor representation of $SO(6)$.

\def\thesubsection{A.3}
\subsection{ Type IIA  $AdS_4\times CP^3$ superspace}

The type IIA superspace whose bosonic body is $AdS_4\times CP^3$ is
parametrized by 10 bosonic coordinates $X^M=(x^{  m},\,y^{m'})$ and
32-fermionic coordinates $\Theta^{\underline\mu}=(\Theta^{\mu\mu'})$
($\mu=1,2,3,4;\,\mu'=1,\cdots,8$). These  combine into the
superspace supercoordinates $Z^{\cal M}=(x^{
m},\,y^{m'},\,\Theta^{\mu\mu'})$. The type IIA supervielbeins are
\begin{equation}\label{IIAsv}
{\mathcal E}^{\mathcal A}=dZ^{\mathcal M}\,{\mathcal E}_{\mathcal
M}{}^{\mathcal A}(Z)=({\mathcal E}^{A},\,{\mathcal
E}^{\underline\alpha})\,,\qquad {\mathcal E}^{A}(Z)=({\mathcal E}^{
a},\,{\mathcal E}^{a'})\,,\qquad {\mathcal
E}^{\underline\alpha}(Z)={\mathcal E}^{\alpha\alpha'}\,.
\ee

\def\thesubsection{A.4}
\subsection{Superspace constraints}
In our conventions the superspace constraint on the bosonic part of
the torsion is
\be\label{torsion}
T^A=-i\mathcal E\Gamma^A\mathcal E+i\mathcal E^A\,\mathcal
E\lambda+\frac{1}{3}{\mathcal E}^A\,\mathcal E^B\nabla_B\phi\,,
\ee
while the constraints on the RR and NS--NS field strengths are
\begin{eqnarray}
F_2&=&-i\,e^{-\phi}\,\mathcal E\Gamma_{11}\mathcal E
+2i\,e^{-\phi}\,\mathcal E^A\,\mathcal E\Gamma_A\Gamma_{11}\lambda+\frac{1}{2}\mathcal E^B\mathcal E^A\,F_{AB}\,,\\
F_4&=&-\frac{i}{2}\,e^{-\phi}\,{\mathcal E}^B{\mathcal
E}^A\,\mathcal E\Gamma_{AB}\mathcal E +\frac{1}{4!}{\mathcal
E}^D{\mathcal E}^C{\mathcal E}^B{\mathcal E}^A\,F_{ABCD}
\,,\\
H_3&=& -i{\mathcal E}^A\,\mathcal E\Gamma_A\Gamma_{11}\mathcal E
+i{\mathcal E}^B{\mathcal E}^A\,\mathcal
E\Gamma_{AB}\Gamma^{11}\lambda +\frac{1}{3!}{\mathcal E}^C{\mathcal
E}^B{\mathcal E}^A\,H_{ABC}\,.
\end{eqnarray}
These differ from the conventional string frame constraints by the
dilatino $\lambda$--term in $T^A$ and related terms in $F_2$, $F_4$
and $H_3$. This is a consequence of the dimensional reduction from
eleven dimensions. The constraints can be brought to a more
conventional string--frame form by shifting the fermionic
supervielbein $\mathcal E^{\underline\alpha}$ by
$-\frac{1}{2}\mathcal E^A(\Gamma_A\lambda)^{\underline\alpha}$
accompanied by a related shift in the connection. Note also that the
purely bosonic part of the torsion, the last term in
\eqref{torsion}, can be eliminated by a proper redefinition of the
spin connection.
\\
\\
{\bf The $D=10$ gamma--matrices $\Gamma^A$} are given by
\bee\label{Gamma10}
&\{\Gamma^A,\,\Gamma^B\}=2\eta^{AB},\qquad
\Gamma^{A}=(\Gamma^{  a},\,\Gamma^{a'})\,,\nonumber\\
&\\
&\Gamma^{  a}=\gamma^{  a}\,\otimes\,{\bf 1},\qquad
\Gamma^{a'}=\gamma^5\,\otimes\,\gamma^{a'},\qquad
\Gamma^{11}=\gamma^5\,\otimes\,\gamma^7,\qquad a=0,1,2,3;\quad
a'=1,\cdots,6\,. \nonumber
\eee
The charge conjugation matrix is ${\mathcal C}=C\otimes C'$.

The fermionic variables $\Theta^{\underline\alpha}$ of IIA
supergravity carrying 32--component spinor indices of $Spin(1,9)$,
in the $AdS_4\times CP^3$ background and for the above choice of the
$D=10$ gamma--matrices, naturally split into 4--dimensional
$Spin(1,3)$ indices and 8--dimensional spinor indices of $Spin(6)$,
i.e. $\Theta^{\underline\alpha}=\Theta^{\alpha\alpha'}$
($\alpha=1,2,3,4$; $\alpha'=1,\cdots,8$).

\def\thesubsection{A.5}
\subsection{$24+8$ splitting of $32$ $\Theta$}

24 of  $\Theta^{\underline\alpha}=\Theta^{\alpha\alpha'}$ correspond
to the unbroken supersymmetries of the $AdS_4\times CP^3$
background. They are singled out by a projector introduced in
\cite{Nilsson:1984bj} which is constructed using the $CP^3$ K\"ahler
form $J_{a'b'}$ and seven $8\times 8$ antisymmetric gamma--matrices
(\ref{gammaa'}). The $8\times 8$ projector matrix has the following
form
\be\label{p6}
{\mathcal P}_{6}=\frac{1}{8}(6-J)\,,
\ee
where the $8\times 8$  matrix
\be\label{J}
J=-iJ_{a'b'}\,\gamma^{a'b'}\,\gamma^7 \qquad {\rm such~ that} \qquad
J^2= 4J+12
\ee
has six eigenvalues $-2$ and two eigenvalues $6$, \emph{i.e.} its
diagonalization results in
\be\label{Jdia}
J=\hbox{diag}(-2,-2,-2,-2,-2,-2,6,6)\,.
\ee
Therefore, the projector (\ref{p6}) when acting on an 8--dimensional
spinor annihilates 2 and leaves 6 of its components, while the
complementary projector
\be\label{p2}
{\mathcal P}_{2}=\frac{1}{8}(2+J)\,,\qquad
\mathcal{P}_2+\mathcal{P}_6=\mathbf 1
\ee
annihilates 6 and leaves 2 spinor components.

Thus the spinor
\be\label{24}
\vartheta^{\alpha\alpha'}=({\mathcal P}_6\,\Theta)^{\alpha\alpha'} \qquad \Longleftrightarrow \qquad
\vartheta^{\alpha a'}\, \qquad a'=1,\cdots, 6
\ee
has 24 non--zero components and the spinor
\be\label{8}
\upsilon^{\alpha\alpha'}=({\mathcal P}_2\,\Theta)^{\alpha\alpha'}\qquad \Longleftrightarrow \qquad
\upsilon^{\alpha i}\, \qquad i=1,2
\ee
has 8 non--zero components. The latter corresponds to the eight
supersymmetries broken by the $AdS_4\times CP^3$ background.

To avoid confusion, let us note that the index $a'$ on spinors is
different from the same index on bosonic quantities. They are
related by the usual relation between vector and spinor
representations, \emph{i.e.} given two $Spin(6)$ spinors
$\psi_1^{\alpha'}$ and $\psi_2^{\alpha'}$, projected as in
(\ref{24}), their bilinear combination $v^{a'}=\psi_1\mathcal
P_6\gamma^{a'}\mathcal P_6\psi_2=\psi_1^{b'}(\mathcal
P_6\gamma^{a'}\mathcal P_6)_{b'c'}\psi_2^{c'}$ transforms as a
6--dimensional 'vector'.

\def\thesubsection{A.6}
\subsection{The explicit form of the geometry of the $AdS_4\times CP^3$ superspace}
The supervielbeins have the following form
\be\label{simplA}
\begin{aligned}
{\mathcal
E}^{a'}(x,y,\vartheta,\upsilon)&=e^{\frac{1}{3}\phi(\upsilon)}\,\left(E^{a'}(x,y,\vartheta)+2i\upsilon\,\frac{\sinh m}
{m}\gamma^{a'}\gamma^5\,E(x,y,\vartheta)\right) \,,
\\
\\
{\mathcal E}^{  a}(x,y,\vartheta,\upsilon) &=
e^{\frac{1}{3}\phi(\upsilon)}\,\left(E^{
b}(x,y,\vartheta)+4i\upsilon\gamma^{  b}\,\frac{\sinh^2{{\mathcal M}/
2}}{{\mathcal M}^2}\,D\upsilon\right)\Lambda_{  b}{}^{
a}(\upsilon)
\\
&{}
\hskip+1cm -e^{-\frac{1}{3}\phi(\upsilon)}\,\frac{R^2}{kl_p}\left(A(x,y,\vartheta)-\frac{4}{R}\upsilon\,\ve\gamma^5\,\frac{\sinh^2{{\mathcal
M}/2}}{{\mathcal M}^2}\,D\upsilon\right) E_7{}^{ a}(\upsilon)\,,
\\
\\
{\mathcal E}^{\alpha i}(x,y,\vartheta,\upsilon) &=
e^{\frac{1}{6}\phi(\upsilon)}\,\left(\frac{\sinh{\mathcal
M}}{{\mathcal M}}\,D\upsilon\right)^{\beta j}\,S_{\beta
j}{}^{\alpha i}\,(\upsilon) -ie^{\phi(\upsilon)}{\mathcal
A}_1(x,y,\vartheta,\upsilon)\,(\gamma^5\varepsilon\lambda(\upsilon))^{\alpha
i}\,,
\\
\\
{\mathcal E}^{\alpha a'}(x,y,\vartheta,\upsilon) &=
e^{\frac{1}{6}\phi(\upsilon)}\,E^{\gamma b'}(x,y,\vartheta)\,\left(
\delta_{\gamma}{}^{\beta}-\frac{8}{R}\,\left(\gamma^5\,\upsilon\,\frac{\sinh^2{{m}/2}}{m^2}\right)_{\gamma
i}\upsilon^{\beta i} \right)S_{\beta b'}{}^{\alpha
a'}\,(\upsilon)\,.
\end{aligned}
\ee
The new objects appearing in these expressions, $m$, $\mathcal M$,
$\Lambda_{  a}{}^{  b}$, $E_7{}^{  a}$ and
$S_{\underline\alpha}^{\underline\beta}$, are functions of $\ups$
and their explicit forms are given in Appendix A.8 while the dilaton
$\phi$, dilatino $\lambda$ and RR one--form $\mathcal A_1$ are given
below. Contracted spinor indices have been suppressed, \emph{e.g.}
$(\ups\ve\gamma^5)_{\alpha i}=\ups^{\beta
j}\ve_{ji}\gamma^5_{\beta\alpha}$, where
$\varepsilon_{ij}=-\varepsilon_{ji}$, $\varepsilon_{12}=1$ is the
$SO(2)$ invariant tensor. Note that $\varepsilon=-i{\mathcal
P_2}\gamma_7{\mathcal P_2}$.

To avoid confusion, let us stress that the indices $i,j$ carried by
the $\mathcal P_2$--projected spinors $\upsilon^{\alpha i}$ and by
the associated quantities, are different from the indices $i,j$ used
in the main text of the paper to define the tangent--space indices
of $S^2$. The indices carried by $\upsilon$ never appear in the main
text. Using the same indices but for labeling different quantities
in some instances which do not cause the confusion we thus avoid
redundant proliferation of different labels.

The covariant derivative is defined as
\bee\label{D}
D\upsilon=\left(d+\frac{i}{R}E^{ a}(x,y,\vartheta)\,\gamma^5\gamma_{
a}-\frac{1}{4}\Omega^{  a
  b}(x,y,\vartheta)\,\gamma_{  a  b}\right)\upsilon \,.
\eee
The type IIA RR one--form gauge superfield is
\be\label{simplB}
\begin{aligned}
{\mathcal A}_1(x,y,\vartheta,\upsilon) &=
R\,e^{-\frac{4}{3}\phi(\upsilon)}\,\left[
\left(A(x,y,\vartheta)-\frac{4}{R}\upsilon\,\ve\gamma^5\,\frac{\sinh^2{{\mathcal
M}/2}}{{\mathcal
M}^2}\,D\upsilon\right)\frac{R}{kl_p}\,\Phi(\upsilon)
\right.\\
&\left.\hspace{40pt}+\frac{1}{kl_p}\left(E^{
a}(x,y,\vartheta)+4i\upsilon\gamma^{  a}\,\frac{\sinh^2{{\mathcal
M}/2}}{{\mathcal M}^2}\,D\upsilon\right)E_{7  a}(\upsilon)
\right]\,.
\end{aligned}
\ee

The RR four-form and the NS--NS three-form superfield strengths are
given by
\be\label{f4h3}
\begin{aligned}
F_4&=d{\mathcal A}_3-{\mathcal A}_1\,H_3=-\frac{1}{4!}{\mathcal E}^{
d}{\mathcal E}^{  c}{\mathcal E}^{  b}{\mathcal E}^{
a}\left(\frac{6}{kl_p}\,e^{-2\phi}\Phi\ve_{  a  b  c  d}\right)
-\frac{i}{2}{\mathcal E}^{B}{\mathcal E}^{A}{\mathcal
E}^{\underline\beta}
{\mathcal E}^{\underline\alpha}e^{-\phi}(\Gamma_{AB})_{\underline{\alpha\beta}}\,,\\
H_3&=dB_2=-\frac{1}{3!}{\mathcal E}^{  c}{\mathcal E}^{ b}{\mathcal
E}^{  a}(\frac{6}{kl_p}e^{-\phi}\ve_{  a  b  c  d}E_7{}^{  d})
-i{\mathcal E}^{A}{\mathcal E}^{\underline\beta}{\mathcal
E}^{\underline\alpha}(\Gamma_A\Gamma_{11})_{\underline{\alpha\beta}}
+i{\mathcal E}^{B}{\mathcal E}^{A}{\mathcal
E}^{\underline\alpha}(\Gamma_{AB}\Gamma^{11}\lambda)_{\underline\alpha}
\end{aligned}
\ee
and the corresponding gauge potentials are
\be\label{B2}
B_2=b_2+\int_0^1\,dt\,i_\Theta H_3(x,y,t\Theta)\,,\qquad \Theta=(\vartheta,\upsilon)\,\\
\ee
\be\label{A3}
\hskip+1.9cm{\mathcal
A}_3=a_3+\int_0^1\,dt\,i_\Theta\left(F_4+\mathcal{A}_1H_3\right)(x,y,t\Theta)\,,
\ee
where $b_2$ and $a_3$ are the purely bosonic parts of the gauge
potentials and $i_\Theta$ means the inner product with
$\Theta^{\underline\alpha}$. Note that $b_2$ is  pure gauge in the
$AdS_4\times CP^3$ solution while $a_3$ is the RR three-form
potential of the bosonic background.

 The dilaton superfield $\phi(\upsilon)$, which depends only on
the eight fermionic coordinates corresponding to the broken
supersymmetries, has the following form in terms of $E_7{}^{
a}(\upsilon)$ and $\Phi(\upsilon)$
\be\label{dilaton1}
e^{\frac{2}{3}\phi(\upsilon)}=\frac{R}{kl_p}\,\sqrt{\Phi^2+E_7{}^{
a}\,E_7{}^{  b}\,\eta_{  a  b}}\,.
\ee
The value of the dilaton at $\upsilon=0$ is
\begin{equation}
e^{\frac{2}{3}\phi(\upsilon)}|_{\upsilon=0}=e^{\frac{2}{3}\phi_0}=\frac{R}{kl_p}\,.
\end{equation}
The fermionic field $\lambda^{\alpha i}(\upsilon)$ describes the
non--zero components of the dilatino superfield and is given by the
equation \cite{Howe:2004ib}
\be\label{dilatino1}
\lambda_{\alpha i}=-\frac{i}{3}D_{\alpha i}\,\phi(\upsilon)\,.
\ee
In the above expressions $E^{  a}(x,y,\vartheta)$,  $E^{
a'}(x,y,\vartheta)$ and $\Omega^{  a  b}(x,y,\vartheta)$ are the
supervielbeins and the $AdS_4$ part of the spin connection of the
supercoset $OSp(6|4)/U(3)\times SO(1,3)$ and $A(x,y,\vartheta)$ is
the corresponding type IIA RR one--form gauge superfield whose
explicit form is given below.

\def\thesubsection{A.7}
\subsection{$OSp(6|4)/U(3)\times SO(1,3)$ supercoset realization
and other ingredients of the $(10|32)$--dimensional $AdS_4\times
CP^3$ superspace}
\setcounter{equation}0

The supervielbeins and the superconnections of the
$OSp(6|4)/U(3)\times SO(1,3)$ supercoset which appear in the
definition of the geometric and gauge quantities of the $AdS_4\times
CP^3$ superspace  can be parametrized in the following way
\be\label{cartan24}
\begin{aligned}
E^{  a}&=e^{  a}(x)+4i\vartheta\gamma^{  a}\,\frac{\sinh^2{{\mathcal
M}_{24}/ 2}}{{\mathcal M}^2_{24}}\,
D_{24}\vartheta,\\
E^{a'}&=e^{a'}(y)+4i\vartheta\gamma^{a'}\gamma^5\,\frac{\sinh^2{{\mathcal
M}_{24}/2}}{{\mathcal M}_{24}^2}\,D_{24}\vartheta\,,
\\
E^{\alpha a'}&=\left(\frac{\sinh{\mathcal M}_{24}}{{\mathcal
M}_{24}}D_{24}\vartheta\right)^{\alpha a'},\\
\Omega^{  a  b}&=\omega^{  a  b}(x)+\frac{8}{R}\vartheta\gamma^{ab}\gamma^5\,
\frac{\sinh^2{{\mathcal M}_{24}/2}}{{\mathcal M}_{24}^2}D_{24}\vartheta\,,\\
\Omega^{a'b'}&=\omega^{a'b'}(y)-\frac{4}{R}
\vartheta(\gamma^{a'b'}-iJ^{a'b'}\gamma^7)\gamma^5\,\frac{\sinh^2{{\mathcal M}_{24}/2}}
{{\mathcal M}_{24}^2}\,D_{24}\vartheta\,,\\
A&=\frac{1}{8}J_{a'b'}\Omega^{a'b'}=A(y)+\frac{4i}{R}\,\vartheta\gamma^7\gamma^5\,\frac{\sinh^2{{\mathcal
M}_{24}/2}}{{\mathcal M}_{24}^2}\,D_{24}\vartheta\,,
\end{aligned}
\ee
where
\be\label{M24}
R\,({\mathcal M}_{24}^2)^{\alpha a'}{}_{\beta b'}=
4\vartheta^{\alpha}_ {b'}\,(\vartheta^{a'}\gamma^5)_\beta
-4\delta^{a'}_{b'}\vartheta^{\alpha c'}(\vartheta\gamma^5)_{\beta
c'} -2(\gamma^5\gamma^{  a}\vartheta)^{\alpha a'}(\vartheta\gamma_{
a})_{\beta b'} -(\gamma^{  a  b}\vartheta)^{\alpha
a'}(\vartheta\gamma_{  a  b}\gamma^5)_{\beta b'}\,.
\ee
The derivative appearing in the above equations is defined as
\be\label{D24}
D_{24}\vartheta={\mathcal P_6}\,(d +\frac{i}{R}\,e^{
a}\gamma^5\gamma_{  a}
+\frac{i}{R}e^{a'}\gamma_{a'}-\frac{1}{4}\omega^{  a b}\gamma_{ a
b} -\frac{1}{4}\omega^{a'b'}\gamma_{a'b'})\vartheta\,,
\ee
where $e^{  a}(x)$, $e^{a'}(y)$, $\omega^{  a  b}(x)$,
$\omega^{a'b'}(y)$ and $A(y)$ are the vielbeins and connections of
the bosonic solution. The $U(3)$--connection $\Omega^{a'b'}$
satisfies the condition
\begin{equation}
{(P^{-})_{a'b'}}^{c'd'}\Omega_{c'd'}=\frac{1}{2}\,({\delta_{[a'}}^{c'}\,{\delta_{b']}}^{d'}\,-\,
{J_{[a'}}^{c'}\,{J_{b']}}^{d'})\Omega_{c'd'}=0\,,
\end{equation}
where $J_{a'b'}$ is the K\"ahler form on $CP^3$.

Let us stress once again that the index $a'$ on spinors is different
from the same index on bosonic quantities. See the end of Appendix
A.5 for more explanation.

\def\thesubsection{A.8}
\subsection{Other quantities appearing in the definition of the
$AdS_4\times CP^3$ superspace}

\be\label{M}
R\,({\mathcal M}^2)^{\alpha i}{}_{\beta j}= 4(\ve\upsilon)^{\alpha
i}(\ups\ve\gamma^5)_{\beta j} -2(\gamma^5\gamma^{
a}\upsilon)^{\alpha i}(\ups\gamma_{  a})_{\beta j} -(\gamma^{ a
b}\upsilon)^{\alpha i}(\ups\gamma_{  a  b}\gamma^5)_{\beta j}\,,
\ee
\be
(m^2)^{ij}=-\frac{4}{R}\upsilon^i\,\gamma^5\,\upsilon^j\,,
\ee

\be
\begin{aligned}
\Lambda_{  a}{}^{  b}&=
\delta_{  a}{}^{  b}-\frac{R^2}{k^2l_p^2}\,\cdot\,
\frac{e^{-\frac{2}{3}\phi}}{e^{\frac{2}{3}\phi}
+\frac{R}{kl_p}\,\Phi}\,{E_{7  a}}\,E_7{}^{  b}\,,
\\
\\
S_{\underline\beta}{}^{\underline\alpha}&=
\frac{e^{-\frac{1}{3}\phi}}{\sqrt2}\left(\sqrt{e^{\frac{2}{3}\phi}
+\frac{R}{kl_p}\,\Phi}-\frac{R}{kl_p}\,
\frac{E_7{}^{  a}\,\Gamma_{  a}\Gamma_{11}}{\sqrt{e^{\frac{2}{3}\phi}
+\frac{R}{kl_p}\,\Phi}}
\,\right)_{\underline\beta}{}^{\underline\alpha}
\end{aligned}
\ee

\be\label{phiE7}\begin{aligned}
E_7{}^{  a}(\upsilon)&=-\frac{8i}{R}\,\upsilon\gamma^{
a}\,\frac{\sinh^2{{\mathcal M}/ 2}}{{\mathcal
M}^2}\,\varepsilon\,{\upsilon}\,,
\\
\Phi(\upsilon)&= 1+\frac{8}{R}\,\upsilon\,\ve\gamma^5\,\frac{\sinh^2{{\mathcal
M}/2}}{{\mathcal M}^2}\,\ve\upsilon\,.
\end{aligned}
\ee
Let us emphasise that the $SO(2)$ indices $i,j=1,2$ are raised and
lowered with the unit matrices $\delta^{ij}$ and $\delta_{ij}$ so
that there is actually no difference between the upper and the lower
$SO(2)$ indices, $\varepsilon_{ij}=-\varepsilon_{ji}$,
$\varepsilon^{ij}=-\varepsilon^{ji}$ and
$\varepsilon^{12}=\varepsilon_{12}=1$.

\def\thesection{}
\def\theequation{B.\arabic{equation}}\label{B}
\section{Appendix B. $CP^3$ Geometry}
\setcounter{equation}0

The Fubini-Study metric on $CP^3$ is
\begin{equation}\label{B1}
ds^2=\rho^{-2}d\bar\zeta_ad\zeta^a-\rho^{-4}\zeta^ad\bar\zeta_a\,\bar\zeta_bd\zeta^b\,,
\end{equation}
where $\zeta^a$ are three complex numbers and
$\rho^2=1+\zeta^a\bar\zeta_a$. Real coordinates adapted to the
$U(3)$ isotropy group can be introduced as follows
\cite{Pope:1984bd}
\begin{eqnarray}
\zeta^1&=&\tan\frac{\theta}{2}\,\sin\alpha\,\sin\frac{\vartheta}{2}\,e^{i(\psi-\chi)/2}e^{i\varphi}\nonumber\\
\zeta^2&=&\tan\frac{\theta}{2}\,\cos\alpha\,e^{i\varphi}\nonumber\\
\zeta^3&=&\tan\frac{\theta}{2}\,\sin\alpha\,\cos\frac{\vartheta}{2}\,e^{i(\psi+\chi)/2}e^{i\varphi}\,,
\end{eqnarray}
where $0\leq\theta,\vartheta\leq\pi$, $0\leq\varphi,\chi\leq2\pi$,
$0\leq\alpha\leq\frac{\pi}{2}$ and $0\leq\psi\leq4\pi$. In these
coordinates the metric becomes
\begin{equation}
ds^2=
\frac{1}{4}\big(d\theta^2+\sin^2\theta(d\varphi+\frac{1}{2}\sin^2\alpha\,\sigma_3)^2\big)
+\sin^2\frac{\theta}{2}\,d\alpha^2
+\frac{1}{4}\sin^2\frac{\theta}{2}\,\sin^2\alpha(\sigma_1^2+\sigma_2^2+\cos^2\alpha\,\sigma_3^2)\,,
\end{equation}
where
\begin{eqnarray}
\sigma_1&=&\sin\psi\,d\vartheta-\cos\psi\,\sin\vartheta\,d\chi\nonumber\\
\sigma_2&=&\cos\psi\,d\vartheta+\sin\psi\,\sin\vartheta\,d\chi\nonumber\\
\sigma_3&=&d\psi+\cos\vartheta\,d\chi
\end{eqnarray}
are three left-invariant one-forms on $SU(2)$ obeying
$d\sigma_1=-\sigma_2\sigma_3$ etc. Notice that with this choice of
coordinates $\theta$ and $\varphi$ parameterize a two-sphere of radius
$\frac{1}{2}$. This two-sphere is topologically non-trivial and
associated to the K\"ahler form on $CP^3$. We choose the $CP^3$
vielbeins as follows
\begin{eqnarray}
e^1&=&\frac{1}{2}d\theta\nonumber\\
e^2&=&\frac{1}{2}\sin\theta(d\varphi+\frac{1}{2}\sin^2\alpha\,\sigma_3)\nonumber\\
e^3&=&-\frac{1}{2}\sin\frac{\theta}{2}\,\sin\alpha\,\sigma_2\nonumber\\
e^4&=&\frac{1}{2}\sin\frac{\theta}{2}\,\sin\alpha\,\sigma_1\nonumber\\
e^5&=&\sin\frac{\theta}{2}\,d\alpha\nonumber\\
e^6&=&\frac{1}{4}\sin\frac{\theta}{2}\,\sin(2\alpha)\,\sigma_3\,.
\end{eqnarray}
Using the fact that
\begin{eqnarray}
de^1&=&0\nonumber\\
de^2&=&2\cot\theta\,e^2\,e^1+2\cot\frac{\theta}{2}\,e^6\,e^5+2\cot\frac{\theta}{2}\,e^4\,e^3\nonumber\\
de^3&=&\cot\frac{\theta}{2}\,e^3\,e^1+\frac{\cot\alpha}{\sin\frac{\theta}{2}}\,e^3\,e^5+\frac{4}{\sin\frac{\theta}{2}\,\sin(2\alpha)}\,e^6\,e^4 \nonumber\\
de^4&=&\cot\frac{\theta}{2}\,e^4\,e^1+\frac{\cot\alpha}{\sin\frac{\theta}{2}}\,e^4\,e^5+\frac{4}{\sin\frac{\theta}{2}\,\sin(2\alpha)}\,e^3\,e^6
\nonumber\\
de^5&=&\cot\frac{\theta}{2}\,e^5\,e^1\nonumber\\
de^6&=&\cot\frac{\theta}{2}\,e^6\,e^1+\frac{2\cot(2\alpha)}{\sin\frac{\theta}{2}}\,e^6\,e^5+\frac{2\cot\alpha}{\sin\frac{\theta}{2}}\,e^4\,e^3
\end{eqnarray}
one can show that the connection can be taken to be
\begin{equation}
\begin{array}{rclcrcl}
\omega^{12}&=&2\cot\theta\,e^2 &\qquad& \omega^{34}&=&\cot\frac{\theta}{2}\,e^2-2\frac{1+\sin^2\alpha}{\sin\frac{\theta}{2}\sin(2\alpha)}\,e^6\\
\omega^{1\tilde a}&=&\cot\frac{\theta}{2}\,e^{\tilde a} && \omega^{35}&=&-\frac{\cot\alpha}{\sin\frac{\theta}{2}}\,e^3\\
\omega^{23}&=&-\cot\frac{\theta}{2}\,e^4 && \omega^{36}&=&\frac{\cot\alpha}{\sin\frac{\theta}{2}}\,e^4\\
\omega^{24}&=&\cot\frac{\theta}{2}\,e^3 && \omega^{45}&=&-\frac{\cot\alpha}{\sin\frac{\theta}{2}}\,e^4\\
\omega^{25}&=&-\cot\frac{\theta}{2}\,e^6 && \omega^{46}&=&-\frac{\cot\alpha}{\sin\frac{\theta}{2}}\,e^3\\
\omega^{26}&=&\cot\frac{\theta}{2}\,e^5 && \omega^{56}&=&\cot\frac{\theta}{2}\,e^2+2\frac{\cot(2\alpha)}{\sin\frac{\theta}{2}}\,e^6\,,
\end{array}
\end{equation}
where $\tilde a=3,4,5,6$. The curvature of $CP^3$ is
\begin{equation}
R^{a'b'}=d\omega^{a'b'}+\omega^{a'}{}_{c'}\omega^{c'b'}=(\delta^{a'}_{c'}\delta_{d'}^{b'}+J_{c'}{}^{a'}J_{d'}{}^{b'})\,e^{c'}e^{d'}+J^{a'b'}J_{c'd'}\,e^{c'}e^{d'}\,,
\end{equation}
where $J_{a'b'}$ are the components of the K\"ahler form with
$J_{12}=J_{34}=J_{56}=1$.

The $U(1)$ part of the connection is
\begin{equation}
A=\frac{1}{8}J_{a'b'}\omega^{a'b'}
=\frac{1}{2}(\cot\theta\,e^2+\cot\frac{\theta}{2}\,e^2-3\frac{\tan\alpha}{2\sin\frac{\theta}{2}}\,e^6)
=\cot\theta\,e^2+\frac{1}{4}d\varphi-\frac{\tan\alpha}{2\sin\frac{\theta}{2}}\,e^6\,,
\end{equation}
and it is easy to verify that it's derivative is proportional to the K\"ahler form
\begin{equation}
dA=2\,e^1\,e^2+2\,e^3\,e^4+2\,e^5\,e^6\,.
\end{equation}


\providecommand{\href}[2]{#2}\begingroup\raggedright\endgroup



\end{document}